\documentclass[twocolumn,pre,showpacs,eqsecnum,floatfix]{revtex4}
\usepackage{graphicx}
\usepackage{amsmath}
\usepackage{bm}
\usepackage{epsfig}
\usepackage{amsfonts}
\usepackage{amssymb}
\usepackage[usenames]{color}
\usepackage[latin1]{inputenc}
\begin{document}

\newcommand{\brm}[1]{\bm{{\rm #1}}}
\newcommand{\Ochange}[1]{{\color{red}{#1}}}
\newcommand{\Ocomment}[1]{{\color{PineGreen}{#1}}}
\newcommand{\Hcomment}[1]{{\color{ProcessBlue}{#1}}}
\newcommand{\Hchange}[1]{{\color{BurntOrange}{#1}}}

\title{Renormalized field theory of collapsing directed randomly branched polymers}
\author{Hans-Karl Janssen}
\author{Frank Wevelsiep}
\affiliation{Institut f\"ur Theoretische Physik III, Heinrich-Heine-Universit\"at, 40225
D\"usseldorf, Germany}
\author{Olaf Stenull}
\affiliation{Department of Physics and Astronomy, University of Pennsylvania, Philadelphia,
PA 19104, USA}
\date{\today}

\begin{abstract}
We present a dynamical field theory for directed randomly branched polymers and in particular their collapse transition. We develop a phenomenological model in the form of a stochastic response functional that allows us to address several interesting problems such as the scaling behavior of the swollen phase and the collapse transition. For the swollen phase, we find that by choosing model parameters appropriately, our stochastic functional reduces to the one describing the relaxation dynamics near the Yang-Lee singularity edge. This corroborates that the scaling behavior of swollen branched polymers is governed by the Yang-Lee universality class as has been known for a long time. The main focus of our paper lies on the collapse transition of directed branched polymers. We show to arbitrary order in renormalized perturbation theory with $\varepsilon$-expansion that this transition belongs to the same universality class as directed percolation.
\end{abstract}
\maketitle

\section{Introduction}

It is well known that a single linear (non-branched) polymer consisting of $N$ monomers in a diluted solution with a good solvent where it is in a swollen conformation with gyration (Flory) radius $R_{N}\sim N^{\nu_{SAW}}$, $\nu_{SAW}\geq1/2$, undergoes a phase transition to a collapsed \textquotedblleft globule\textquotedblright\ with Flory radius $R_{N}\sim N^{1/d}$ (where $d$ is
the dimensionality of space) if the solvent quality deteriorates as it usually does when the temperature of the solution is lowered below the so-called $\theta$-point. Many theoretical tools have been successfully applied to linear polymers, and considerable progress has been made in understanding this collapse transition~\cite{Sch99}.

In contrast, much less is known about collapse transition of randomly branched polymers, which are usually assumed being macroscopically either isotropic or directed. As far as the isotropic case is concerned, a number of numerical studies has been performed over the last 20 years or so~\cite{HsGr05,DeHe83,HeSe96,SeVa94,FGSW92/94,JvR97/99/00}.
The picture that arises from these studies is much more complex than that for the collapse of linear polymers. The basic reason for this added complexity is the possibility
for introducing more than one fugacity to drive the collapse. It
appears that the line of collapse transitions in the phase diagram consist of 2 qualitatively different parts. One part of the line, called the line of $\theta$-transitions, corresponds to
continuous transitions with universal critical exponents from swollen polymer
configurations with tree-like character to compact coil-like configurations.
The other part of the transition line, the line of $\theta^{\prime}%
$-transitions, corresponds to the collapse of swollen foam- or sponge-like polymers with
many cycles of bonds between the monomers to vesicle-like compact
structures. If this transition is assumed to be continuous, one finds nonuniversal exponents~\cite{HsGr05}. The two different parts of the collapse-transition
line are separated by a higher multicritical point which belongs to the
isotropic percolation universality class. One of the open questions regarding the phase diagram is the existence of a possible further transition line between the configurations of
the collapsed polymers.

As far as directed branched polymers (DBP) and their collapse (CDBP) is concerned, the theoretical picture~\cite{ReCo82,Dh87,HeSe96,KnVan02,Imb04} is somewhat clearer than in the isotropic case, in particular in $1+1$-dimensions (one transversal and one longitudinal; in the following, we will denote the full space dimension as $D=d+1$). Noting that
branched polymers, lattice animals, and lattice trees belong to the same
universality class, one can use the results of Dhar \cite{Dh87} on collapsing
directed strongly embedded or site animals. This way, Henkel and Seno
\cite{HeSe96} have shown that there is only one type of $\theta$-transitions
which describes the collapse of directed branched polymers in $(D=2)$-dimensions. This collapse transition belongs to the directed percolation (DP) universality class. Furthermore, Dhar has shown in Ref.~\cite{Dh87} that  the collapse transition is also of directed percolation type in $D=3$  if a special relation between the potentials (fugacities) of directed site animals holds. Flory theory was
applied by Redner and Coniglio \cite{ReCo82}. They correctly derived the upper
critical dimension $d_{c}=4$, but they found critical exponents for the
CDBP transition that are different from the DP exponents which, however, is incorrect as we shall see.

Whereas a comprehensive field theory for swollen (extended) polymers has been existing for quite some time, the field theory for the collapse of branched polymers is much less developed. As far as we know, there exist to date no field theory for the directed case. For the isotropic case, there is the seminal work by Lubensky and Isaacson \cite{LuIs78} and Harris and Lubensky \cite{HaLu81}. However, it turns out that these papers, as far as they consider the collapse transition, contain a fundamental error in the renormalization procedure (they overlook a required renormalization), and as a consequence the long-standing $1$-loop results for the collapse transition are not entirely correct~\cite{janssen_stenull_unpub}. In an upshot, one may say that there are important open questions and unresolved issues in the theory for the collapse of directed and isotropic branched polymers.

Here and in a following publication~\cite{janssen_stenull_unpub}, we develop dynamical field theories for the collapse of directed and isotropic branched polymers, respectively. The idea behind these theories is to start out with stochastic activation processes which lead to percolation clusters and then to exploit the well know connection between branched polymers and lattice animals, i.e., cluster of a given mass. We concentrate on large animals below the percolation point, and we equip the theory with enough parameters to allow for triciticality which corresponds to the collapse of large branched polymers. As alluded to above, the directed case has the benefit of being simpler than the isotropic case. It allows us to learn about the fundamental structure of dynamical field theories for the collapse of branched polymers and thereby to sharpen our tools for the more complicated isotropic case. The most important concrete result of the present paper is that we show to arbitrary order in renormalized perturbation theory with $\varepsilon$-expansion that the CDBP transition is of DP-type.

The outline of the remainder of this paper is as follows: In Sec.~\ref{sec:towardsFieldTheory}, we develop our field theoretic model. In Sec.~\ref{sec:meanFieldTheory} we analyze our model in mean-field theory. We determine the mean-field phase diagram of DBP and discuss their scaling behavior at the collapse transition ignoring fluctuations. In Sec.~\ref{sec:nonCollapsing}, we briefly discuss swollen DBP by making contact to established theories. In Sec.~\ref{sec:fieldTheory}, we present the core of our field theoretic analysis of the collapse transition. We discuss the scaling invariances of our model at this transition and their consequences. We calculate the counter-terms required to renormalize our theory in a 1-loop calculation and by using Ward identities. We set up and solve renormalization group equations that provide us with the scaling form of the equation of state and correlation lengths. In Sec.~\ref{sec:polymerScaling}, we translate our field theoretic results for the collapse transition into a form that is more commonly used in polymer theory, and we compare them to numerical simulations. In Sec.~\ref{sec:concludingRemarks}, we give a few concluding remarks. There are tow appendixes. In Appendix~\ref{app:1Loop}, we present some details of our 1-loop calculation, and in Appendix~\ref{app:essentialSingularities}, we discuss essential singularities.

\section{Towards a dynamical field theory of directed branched polymers}
\label{sec:towardsFieldTheory}

\subsection{A generalized directed percolation process}

In this subsection, we develop a field theoretic stochastic functional
\cite{Ja76,DeDo76,Ja92,JaTa05} for DBP based on very general arguments alluding to the universal properties of a corresponding Markoffian
stochastic percolation process. We consider polymers as clusters in $d$ transversal and $1$ longitudinal (time) directions. Below the percolation threshold all of the clusters generated by such a process are finite. The distribution function for the large clusters decays exponentially with their mass (number of monomers). It is well known \cite{StAh92} that untypical very large clusters with linear sizes essentially larger than the correlation lengths are generated as rare events. The statistical properties of these fractal clusters belong to the universality class of lattice animals. Thus, they also form  the prototype model for single randomly branched polymers in a dilute solution. Allowing for a further mechanism in the percolation process which favors contacts of the monomers, we introduce a possible tricritical instability that leads to the collapse of the large clusters describing the branched polymers.

A Langevin equation which describes directed percolation processes allowing for tricritical behavior has been well known for many years \cite{JaTa05,Ja81,Ja01,OhKe87,Ja87,JaMuSt04,Ja05,Lue06,HeHiLu08}. It is based on the fundamental phenomenological principles of absorbing processes in conjunction with a density and
gradient expansion. The basic variable or field is the density $n(\mathbf{x},t)$ of agents (infected individuals), which models the fractal monomer-density of the polymer generated from a given source $\tilde{h}(\mathbf{x},t)$. This choice of field is possible because the stochastic process generates clusters such that the numbers of monomers, branching-, and end-points are all of the same order which is in contrast to the case of linear polymers. In other words, the density is a proper field for the current problem because branched polymers are really ``fur-bearing animals''. The Langevin equation in Ito-interpretation is given by
\begin{subequations}
\label{StochProz}
\begin{equation}
\lambda^{-1}\dot{n}=R(n)\,n+D_{1}(n)\nabla^{2}n+D_{2}(n)\,(\nabla
n)^{2}+\ldots+\tilde{h}+\zeta\,, \label{LangGl}
\end{equation}
where $\lambda$ is a kinetic coefficient, and the noise correlation reads
\begin{equation}
\overline{\zeta(\mathbf{x},t)\zeta(\mathbf{x}^{\prime},t^{\prime})}
=\lambda^{-1}\Big[Q(n)\,n(\mathbf{x},t)+\ldots\Big]\,\delta(\mathbf{x}
-\mathbf{x}^{\prime})\,\delta(t-t^{\prime})\,. \label{LangKorr}
\end{equation}
\end{subequations}
We assume Gaussian noise which we can for our purposes without loss of generality because higher order noise does not change the universal properties of the process. The leading terms of the expansions are
$D_{1}(n)=1+c_{1}n+\ldots$, $D_{2}(n)=-c_{2}+\ldots$, $R(n)=-r-g^{\prime
}n/2-f^{\prime}n^{2}/6+\ldots$, and $Q(n)=g+\ldots$. Further terms in the
expansions are possible but they turn out being irrelevant. In mean-field theory, the
percolation transition occurs at $r=0$. As announced above, we restrict ourself in the following
to the region $r>0$ in which only non-percolating directed branched clusters with a typical linear size $\xi\sim r^{-1/2}$ are generated from a time- and space-localized source. However, our primary interest is in the rare events where the fractal clusters of linear size essentially larger than $r^{-1/2}$ are generated. As long as $g^{\prime}>0$, the second order term $f^{\prime}n^{2}$ of the rate $R$ is irrelevant. We permit both signs of $g^{\prime}$ so that our model accounts for a tricritical instability.
Consequently, we need the second order term with $f^{\prime}>0$ for stabilization
purposes, i.e., to limit the density $n$ to finite values. The parameters
$c_{i}$ (which are irrelevant for the original percolation problem) are assumed to be positive to
smooth density fluctuations. An essential property of percolation processes is the linear dependence of the noise correlation (\ref{LangKorr}) on the density if $n\to 0$ for $g>0$. This property guarantees that the percolation process is really absorbing \cite{Ja05}, and guarantees the description by a density variable. Last but not least, all the terms lead to $\dot{n}=0$ if $n=0$ and $\tilde{h}=0$ on grounds of the absorbing state.

The approach that we are taking focuses on general principles for processes belonging to the same
universality class and is therefore necessarily phenomenological~\cite{JaTa05}. We devise our
field theoretic model representing the universality class using a purely
mesoscopic stochastic formulation based on the correct order parameters
identified through physical insight in the nature of the critical phenomenon.
Hence, the stochastic response functional that we are about to derive stays in full analogy to the
Landau-Ginzburg-Wilson functional and provides a reliable starting point of the field theoretic method.

\begin{figure}[ptb]
\includegraphics[width=4.5cm]{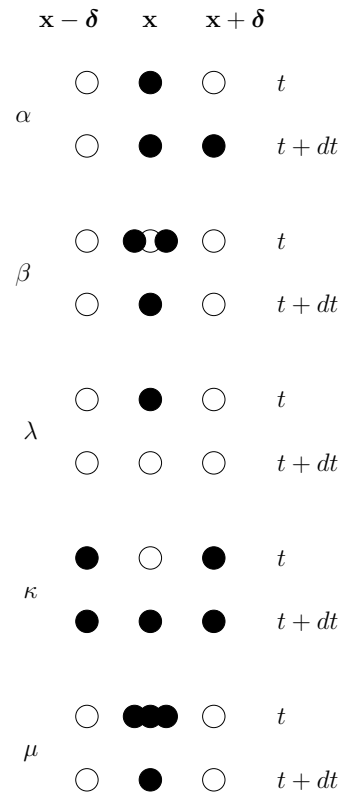}
\caption{Reactions leading to tricritical DP. Open circles indicate lattice sites. Solid dots symbolize agents occupying these sites.}
\label{fig:reactions}
\end{figure}

Alternatively, one might use the so-called \textquotedblleft exact\textquotedblright\ approach which, as a central step, consists of reformulating a microscopic master-equation for chemical reactions as a bosonic field theory on a lattice. For a recent excellent review article on this method see \cite{TaHoVo05}. We choose not to use this method as our main approach because we feel that it treats universal properties not as transparently  as the purely mesoscopic formulation does. Nonetheless, we think that the lattice reactions which would be the starting point for the \textquotedblleft exact\textquotedblright\ approach to the current problem have some pedagogical value as they can help to nurture ones intuition about the epidemic process. Here, these reactions comprise the well known reactions leading to DP:
\begin{subequations}
\label{Reakt}
\begin{align}
X(\mathbf{x})\quad &  \overset{\alpha}{\longrightarrow}\quad X(\mathbf{x}
)+X(\mathbf{x}+\bm{\delta})\,,\label{hin1}\\
2X(\mathbf{x})\quad &  \overset{\beta}{\longrightarrow}\quad X(\mathbf{x}
)\,,\label{back1}\\
X(\mathbf{x})\quad &  \overset{\lambda}{\longrightarrow}\quad\varnothing\,,
\label{back0}
\end{align}
where $\mathbf{x}$ denotes a lattice point, $\bm{\delta}$ denotes a vector to a
neighboring point, and $X(\mathbf{x})$ stands for an agent (particle) at $\mathbf{x}$. Moreover, because we are interested in tricriticality, we need additional reactions that produce more compact clusters:
\begin{align}
X(\mathbf{x}-\bm{\delta})+X(\mathbf{x}+\bm{\delta})\quad   \overset{\kappa
}{\longrightarrow}&\quad X(\mathbf{x}-\bm{\delta })+X(\mathbf{x})
\nonumber\\
&+X(\mathbf{x} +\bm{\delta})\,,
\label{hin2}\\
3X(\mathbf{x})\quad   \overset{\mu}{\longrightarrow}&\quad X(\mathbf{x})\,.
\label{back2}%
\end{align}
\end{subequations}
For illustrations of the reactions defined in Eq.~(\ref{Reakt}), see Fig.~\ref{fig:reactions} and Refs.~\cite{Lue06,Grass06}. Note that the single site back-reactions, Eqs.\ (\ref{back1}) and (\ref{back2}) act against multiple occupations of lattice sites and are thus reminiscent of an excluded-volume interaction.  As mentioned above, we will not use the \textquotedblleft exact\textquotedblright\ approach in this paper. However, for the adherents of the \textquotedblleft exact\textquotedblright\ approach, we mention without presenting further details that we have verified that he bosonic representation of the reaction equations (\ref{Reakt}) leads to the same stochastic functional (called an action in that approach) as the purely mesoscopic approach.

Returning to the latter approach, we now recast the Langevin equations~(\ref{StochProz})  as a stochastic response functional \cite{Ja76,DeDo76,Ja92,JaTa05}
\begin{align}
\mathcal{J}_{\rm{DP}}= &  \int d^{d}xdt\,\lambda\tilde{n}\Big\{\lambda^{-1}
\partial_{t}n+(r-\nabla^{2})n+\frac{g^{\prime}}{2}n^{2}\nonumber\\
&+\frac{f^{\prime}}{6}n^{3}  -c_{1}n\nabla^{2}n+c_{2}(\nabla n)^{2}-\frac{g}{2}\tilde
{n}n\Big\}\,.\label{J_anfang}
\end{align}
Here, we have neglected further higher order terms that will not result in any relevant contributions to our final response functional. $\mathcal{J}_{DP}$ describes the statistics of DP clusters generated
by the stochastic process (\ref{StochProz}). Having this functional, one can calculate the average of any observable that is polynomial in $n$ and $\tilde{n}$, say $\mathcal{O}[n,\tilde{n}]$, by functional integration with weight $\exp(-\mathcal{J}_{\rm{DP}})$,
\begin{align}
\langle\mathcal{O}\rangle_{\rm{DP}} &  = \int \mathcal{D} [n, \tilde{n}] \, \mathcal{O}[n,\tilde{n}] \exp(-\mathcal{J}_{\rm{DP}})\nonumber\\
&=:\operatorname{Tr}\big(\mathcal{O}[n,\tilde{n}] \exp(-\mathcal{J}_{\rm{DP}})\big)\, .
\label{unrestrAv}
\end{align}
Here and in the following, path integrations are  always interpreted in the sense of the (mathematical save) Ito-discretization.
One has to be careful to set the range of the variables of the functional integration correctly as there are some subtleties involved. Some authors \cite{TaHoVo05} use
functional integrals of the type featured in Eq.~(\ref{unrestrAv}) purely in a
restricted sense as a device for generating a perturbation expansion. This
precautionary view arises from a narrow interpretation of the bosonic functional
integral after elimination of irrelevant terms including higher
order monomials of the fields which are in general required to ensure the convergence when fields become large. However these convergence problems can and should be avoided by choosing the support for the functional integration properly, as was shown e.g.\ by Ciafaloni and Onofri
\cite{CiOn79} in the case of Reggeon field theory (RFT which is equivalent to DP). This type of consideration leads here to the rule that $n$ is to be integrated along the positive real axis whereas the integration
of $\tilde{n}$ is performed along the full imaginary axis. Of course, deviations in
finite regions, as e.g.\ suggested by saddle-points, are possible.

One of the most important observables in the current problem is the mass or total number of monomers of a cluster,
\begin{equation}
\mathcal{M}=\lambda\int d^{d}xdt\,n(\mathbf{x},t) \, ,
\end{equation}
where we have included the kinetic coefficient $\lambda$ into the definition for later convenience (if we were not including $\lambda$ in the definition of $\mathcal{M}$, we had to multiply $\mathcal{M}$ in subsequent formulas by $\lambda$ on dimensional grounds).
Because we are mainly interested in the scaling behavior of a single large polymer with $N\gg 1$ monomers, we will focus in the following on averages that are restricted to clusters of a given mass $N$, and we assume without loss of generality that this cluster emanates from a weak source $\tilde{h}(\mathbf{x},t)=q\delta(\mathbf{x})\delta(t)$ of agents at
the origin $\mathbf{x}=0$ at time $t=0$. For our general observable $\mathcal{O}[n]$, this leads to
\begin{align}
\langle\mathcal{O}\rangle_{N}\mathcal{P}(N) &  =\langle\delta
\bigl(N-\mathcal{M}\bigr)\mathcal{O}[n]\exp\bigl(
q\tilde{n}(\mathbf{0},0)\bigr)\rangle_{\rm{DP}}
\nonumber \\
& \simeq q\langle\delta
\bigl(N-\mathcal{M}\bigr)\mathcal{O}[n]
\tilde{n}(\mathbf{0},0))\rangle_{\rm{DP}}  \,,\label{restrAv}
\end{align}
where $\mathcal{P}(N)$ is the probability distribution for finding a cluster
of given mass $N$ with the last equation holding only asymptotically for large $N$ and small $q$. Note that the zeroth-order term in the Taylor expansion leading to the second line of Eq.~(\ref{restrAv}) vanishes because $\langle  \mathcal{F} [n]\rangle_{\rm{DP}} =  \mathcal{F} [0]$ for any functional $\mathcal{F}$ of $n$ because DP is an absorbing process. Note also, that the formalism allows with ease to ask for the probability of clusters
generated by several sources at different points $(\mathbf{r_{i}},t_{i})$ by
inserting more fields $\tilde{n}(\mathbf{r_{i}},t_{i})$ in the exponential of the average
(\ref{restrAv}). Equation~(\ref{restrAv}) implies that the probability distribution for finding a cluster of mass $N$ is given by~\cite{Ja05}
\begin{align}
\mathcal{P}(N) &  =\left\langle\delta\bigl(N-\mathcal{M} \bigr)\exp\bigl(q\tilde{n}(\mathbf{0},0)\bigr)\right\rangle_{\rm{DP}}\nonumber\\
& \simeq q\left\langle\delta\bigl(N-\mathcal{M} \bigr)\tilde{n}(\mathbf{0},0)\right\rangle_{\rm{DP}}\,.
\label{Prob_N}%
\end{align}
Note that $\mathcal{P}(N)$ is correctly normalized, $\int dN\,\mathcal{P}(N)=\langle  \exp (q \tilde{n})\rangle_{\rm{DP}}= 1$, because $\langle  \mathcal{F} [\tilde{n}]\rangle_{\rm{DP}} =  \mathcal{F} [0]$ for any functional $\mathcal{F}$ of $\tilde{n}$ due to causality.

$\mathcal{P}(N)$ is asymptotically proportional (up to nonuniversal amplitudes and an exponential factor $\mu_{0}^{N}$, where $\mu_{0}$ is an effective coordination number of the lattice) to the lattice animal number $\mathcal{A}_{N}$ which plays an important role in percolation theory. $\mathcal{A}_{N}$ measures the number of directed clusters of size $N$, weighted by fugacities for different cluster properties as contacts, loops, and so on. In terms of the number of configurations $\mathcal{C} (N, b , c)$ consisting of $N$ sites, $b$ bonds and $c$ not bounded contacts, $\mathcal{A}_{N}$ can be expressed as~\cite{HaLu81}
\begin{align}
\mathcal{A}_{N} = \sum_{b,c} \mathcal{C} (N, b , c) \Lambda_1^b \Lambda_2^c \, .
\end{align}
The fugacities $\Lambda_1$ and $\Lambda_2$ correspond  to the parameters of the stochastic functional $\mathcal{J}_{\rm{DP}}$. The collapse transition occurs for large $N$ when $\Lambda_2$ becomes critical because the latter rewards contact-rich animals. If we assume universality, the probability distribution $\mathcal{P}(N)$ and the cluster number should be related for $N\gg 1$ via
\begin{align}
\mathcal{P}(N) \simeq \frac{\mathcal{A}_{N} \mu_{0}^{-N}}{\sum_{N^\prime}  \mathcal{A}_{N^\prime} \mu_{0}^{-N^\prime}} \, ,
\end{align}
and conversely
\begin{align}
\mathcal{A}_{N}\simeq\mu_{0}^{N}\mathcal{P}(N)\, ,
\end{align}
a relation that we shall use later on.

Now we turn to the signature of the collapse transition. Chemically, in a diluted solution the transition occurs when the hard core repulsion of the monomers is exactly compensated by their weak effective attraction, i.e., when the  second Virial coefficient $A^{(2)}$ in the expansion of the osmotic pressure in powers of the polymer density vanishes~\cite{Sch99}. Up to a minus sign this second virial-coefficient is proportional to the space integral of the correlation function between two polymers in the solution. Hence, in our formalism, the signature of the collapse-transition is the vanishing of the space-time integral of the two-polymer correlation
\begin{equation}
\label{orderParamDef}
\mathcal{C}^{(2)}(N;\mathbf{x},t)=\mathcal{P}^{(2)}(N;\mathbf{x},t)-\mathcal{P}^{(2)}(N;\infty)\,,
\end{equation}
where
\begin{equation}
\mathcal{P}^{(2)}(N;\mathbf{x},t)\simeq\langle\delta\bigl(N-\mathcal{M}%
\bigr)\,\tilde{n}(\mathbf{x},t)\tilde{n}(0,0)\rangle_{\rm{DP}}\,,
\end{equation}
is asymptotically the probability distribution of two large clusters of total large mass $N$ generated by two roots at $(\mathbf{0},0)$ and $(\mathbf{x},t)$.
If the distance of the point $(\mathbf{x},t)$ from the origin is substantial
bigger than their linear sizes, the two clusters decouple, and we have
\begin{equation}
\mathcal{P}^{(2)}(N;\mathbf{x},t)\simeq\mathcal{P}^{(2)}(N;\infty)=\int
_{0}^{\infty}dN^{\prime}\,\mathcal{P}(N^{\prime})\mathcal{P}(N-N^{\prime})\,,
\end{equation}
i.e., $\mathcal{C}^{(2)}(N;\mathbf{x},t)$ becomes a cumulant. The second virial-coefficient is proportional to the total measure of this correlation times the factor $N^{1/2}$, which comes into play if one considers only clusters of the same mass $N$,
\begin{align}
A^{(2)}&\sim -N^{1/2}\bigl(\mathcal{P}(N)\bigr)^{-1}\lambda\int d^{d}xdt\mathcal{C}^{(2)}(N;\mathbf{x},t)q     \nonumber\\
&\simeq -N^{1/2}\lambda\int d^{d}xdt \langle \tilde{n}(\mathbf{x},t)\tilde{n}(\mathbf{0},0)\rangle^{(\rm{cum})}_{N}\,.
\label{vir2}
\end{align}

Instead of working with quantities discussed above as functions of $N$, it is often more convenient to work with their Laplace-transformed counterparts. For $\mathcal{P}(N)$ we then have the representation by an inverse Laplace transformation
\begin{align}
\mathcal{P}(N)&=\int_{\sigma-i\infty}^{\sigma+i\infty}\frac{dz}{2\pi
i}\,\mathrm{e}^{zN}\,\langle\exp\bigl(-z\mathcal{M+}q\tilde{n}
(0,0)\bigr)\rangle_{\rm{DP}}\nonumber\\
&=\int_{\sigma-i\infty}^{\sigma+i\infty}\frac{dz}{2\pi
i}\,\exp\bigl(zN+q\Phi(z)+O(q^{2})\bigr)
\,, \label{Inv-Lapl}
\end{align}
with $\Phi(z)=\langle\tilde{n}\rangle_{z}$, where
$\langle\cdots\rangle_{z}$ denotes  averages taken with respect to the new stochastic functional
\begin{equation}
\mathcal{J}_{z}=\mathcal{J}_{\rm{DP}}+z\mathcal{M} \, . \label{Jz}
\end{equation}
Note that the normalization condition for $\mathcal{P}(N)$ results in $\Phi(0)=0$. The function $\Phi(z)$ corresponds to the generating function of the animal numbers $\mathcal{A}_N$, and plays in the following the role of an order parameter. Before moving on to mean-field theory, we would like to mention that a shifted version of the Laplace variable $z$ will play later on a role akin to that of an external field.

\subsection{Mean-field theory}
\label{sec:meanFieldTheory}

In this subsection, we discuss various aspects of DBP in mean-field theory. Our motivation for presenting this mean-field theory is twofold. (i) The field theory that we are going to present in the following sections is fairly involved. Mean-field theory, on the other hand, allows us to obtain results with relative ease and may serve the reader as a warm up for the more difficult sections to come. (ii) At mean-field level, various results for DBP are well known. By reproducing these results, we will find that our model satisfies important consistency checks.

Interested in a mean-field approximation in the spirit of Landau theory as the starting point of the systematic perturbation expansion, we look for stationary and homogeneous saddle-points of $\mathcal{J}_{z}$, which are determined by
\begin{subequations}
\begin{align}
\lambda^{-1}\frac{\delta\mathcal{J}_{z}}{\delta n}=&
r\tilde{n}+g^\prime\tilde{n}n+\frac{f^\prime}{2}\tilde{n}n^{2}-\frac{g}{2}\tilde{n}^{2}+z=0\,,\label{SP1}\\
\lambda^{-1}\frac{\delta\mathcal{J}_{z}}{\delta\tilde{n}}=& rn+\frac{g^\prime}{2}n^{2}+\frac{f^\prime}{6}n^{3}-g\tilde{n}n=0 \,. \label{SP2}
\end{align}
\end{subequations}
These equations are solved by
\begin{subequations}
\begin{align}
&\left. n\right\vert_{SP}=0\,,\\
\label{PhiMF}
\Phi(z) :=& \left. \tilde{n}\right\vert_{SP}=(r-\tau(z))/g\\
\tau(z)=&\sqrt{2gz+r^{2}}\,.
\end{align}
\label{SPL}
\end{subequations}
The correct sign of the square root is determined from the normalization condition $\Phi(0)=0$. The equation of state for $\Phi(z)$
shows a branch-point on the negative real axis at $z_{c}=-r^{2}/2g$ where $\tau=0$ becomes critical. Note that the presence of $\left. \tilde{n}\right\vert_{SP}$ in Eq.\ (\ref{SP2}) leads effectively to the replacement of the under-critical parameter $r$ by the critical parameter $\tau$.
Hence, this equation leads to the usual mean-field equation of tricritical DP \cite{JaTa05,OhKe87,Ja87,JaMuSt04,Ja05,Lue06,HeHiLu08}. It follows that the
saddle-point solution, Eqs.\ (\ref{SPL}) is stable as long as $g^\prime \geq 0$. If $g^\prime$ becomes negative, Eq.\ (\ref{SP2}) developes continuously a non-zero positive solution $\left. n\right\vert_{SP}=-3g^{\prime}/f^{\prime}$
at the critical value $\tau=0$, signalling the collapse to a compact cluster as a continuous phase transition.

\begin{figure}[ptb]
\includegraphics[width=5cm]{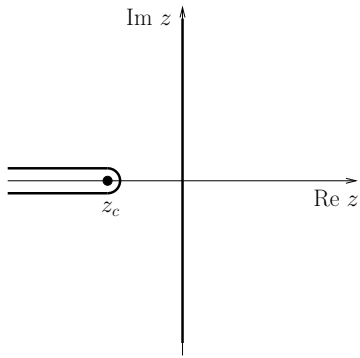}
\caption{Deformation of the contour of integration for the inverse Laplace transformation. Originally, the integration is along the imaginary axis. After deformation, it nestles to and leads around the branch cut that terminates at the branch point $z_c$ on the negative real axis.}
\label{fig:complexInt}
\end{figure}

Qualitatively, it is the branch-point singularity that determines the form of $\mathcal{P}(N)$, and higher orders in perturbation theory, though leading to quantitative improvement, do not result in qualitative changes of that from. To get the asymptotic expansion of $\mathcal{P}(N)$, we deform the contour of the complex integral in Eq.~(\ref{Inv-Lapl}) as illustrated in Fig.~\ref{fig:complexInt}. If $\operatorname{Disc}\Phi$ denotes the discontinuity of the function $\Phi$ at the branch cut, we get
\begin{align}
\mathcal{P}(N)
&\simeq q\mathrm{e}^{z_{c}N+q\Phi(z_{c})}\int_{0}^{\infty}dx\,\frac
{\operatorname*{Disc}\Phi(z_{c}-x)}{2\pi i}\mathrm{e}^{-xN}
\nonumber\\
&\simeq \frac{q\mathrm{e}^{z_{c}N+q\Phi(z_{c})}}{N}\int_{0}^{\infty}dx\,\frac
{\operatorname*{Disc}\Phi(z_{c}-x/N)}{2\pi i}\mathrm{e}^{-x}\,.
\label{Probab_N}
\end{align}
The non-universal factor $q\mathrm{e}^{z_{c}N+q\Phi(z_{c})}$ is common to all the quantities described by Eq.\ (\ref{restrAv}) and cancels therefore in all averages $\langle\mathcal{O}\rangle_{N}$. Note that for large $N$ only the small neighborhood of $z_{c}$ matters.
In mean-field theory, it is easy to calculate $\mathcal{P}(N)$ without further approximation. We obtain
\begin{align}
\label{PNfinal}
\mathcal{P}(N) &=\frac{q}{\sqrt{2\pi g}}N^{-3/2}\exp\Big[-\frac{\bigl(rN-q\bigr)^{2}}
{2gN}\Big]\nonumber\\
&\simeq qN^{-3/2}\exp[-Nr^{2}/2g] \, .
\end{align}
This distribution has its maximum near the characteristic value $N=N_0=q/r$. As stated above, we are, however, mainly interested in the rare events with $N\gg N_0$ where the asymptotic forms are valid. Note that the second line which gives the large-$N$ behavior is in perfect agreement with the well known asymptotic result $\mathcal{P}^{(\text{mf})}(N)\sim N^{-\theta^{(\text{mf})}}\exp(-N/N_{\infty}^{(\text{mf})})$ with the so-called entropic exponent given in mean-field approximation by $\theta^{(\text{mf})}=3/2$ and a non-universal number $N_{\infty}^{(\text{mf})}=2g/r^2$.

Gaussian fluctuations are governed by the second variations of the stochastic functional $\mathcal{J}_z$. After Fourier transformation in space and time, we have
\begin{subequations}
\begin{align}
\Gamma_{1,1}(\mathbf{q},\omega)_{\rm{mf}}  &  =\left.  \frac{\delta\mathcal{J}_{z}
[\tilde{n},n]}{\delta\tilde{n}\delta n}\right\vert
_{SP}=i\omega+\lambda(\tau+\mathbf{q}^{2})\,,
\label{Gamma11}\\
\Gamma_{0,2}(\mathbf{q},\omega)_{\rm{mf}}  &  =\left.  \frac{\delta\mathcal{J}_{z}
[\tilde{n},n]}{\delta n\delta n} \right\vert_{SP} =\lambda \bigl(g^\prime+c\mathbf{q}^{2}\bigr)\Phi\,,
\end{align}
\end{subequations}
where $c=c_1+c_2$. $\Gamma_{2,0;\rm{mf}}=-g\left. n\right\vert_{SP}$ is zero. The propagator follows from the inverse of $\Gamma_{1,1;\rm{mf}}$  in $(\mathbf{q},t)$-representation  as
\begin{equation}
G_{1,1}(\mathbf{q},t;z)= \theta(t)\exp \left[- \lambda \left(\tau(z) + \brm{q}^2\right) t \right]\,,
\label{Propag}
\end{equation}
and the $\tilde{n}$-correlation function integrated over space and time is
\begin{equation}
G_{0,2}(\mathbf{q}=0,\omega=0;z)= -\frac{\Gamma_{0,2}(\mathbf{0},0)_{\rm{mf}}}{\left\vert \Gamma_{1,1}(\mathbf{0},0)_{\rm{mf}}\right\vert^2}\\
=-\frac{g^{\prime}(r-\tau(z))}{\lambda g \tau(z)^2}\,.
\end{equation}
The second virial coefficient follows from Eq.\ (\ref{vir2}) to leading order as
\begin{equation}
A^{(2)} \sim N^{2}g^{\prime}r/g\,.
\end{equation}
It vanishes at the mean-field collapse transition, that is for $g^{\prime}r/g=0$.

Next, we calculate the average monomer density $\langle n (\brm{x}, t) \rangle_N$ emanating from a source at $ (\brm{0}, 0)$. For this calculation, we recall Eq.~(\ref{restrAv}). When specialized  to $\mathcal{O} [n, \tilde{n}] = n (\brm{x}, t)$, the second line of Eq.~(\ref{restrAv}) is equal to the Laplace transform of $\langle n(\mathbf{x},t)\tilde{n}(\mathbf{0},0)\rangle_{z} $ the latter being
the (space-wise) Fourier transform of $G_{1,1}(\mathbf{q},t;z)$, Eq.\ (\ref{Propag}). Collecting, we find that  $\langle n (\brm{q}, t) \rangle_N$ can be written as
\begin{align}
\label{densinter1}
\langle n (\brm{q}, t) \rangle_N \simeq \mathcal{P}(N)^{-1}q \int \frac{dz}{2 \pi i }  \, G_{1,1}(\mathbf{q},t;z) \exp \left( z N \right)\, ,
\end{align}
where the integration path is taken appropriately as outlined above for the probability distribution.
The momentum integration back to the $(\mathbf{x},t)$ representation is straightforward, and we obtain
\begin{align}
\label{densFinal}
\langle n (\brm{x}, t) \rangle_N = \frac{g \lambda t \theta(t)}{(4 \pi \lambda t)^{d/2}} \,  \exp \left[ - \frac{\brm{x}^2}{4 \lambda t} - \frac{g (\lambda t)^2}{2N}\right] .
\end{align}
Form this result, we can read off directly that the longitudinal and transversal radii of gyration scale as $R_{\parallel}\sim N^{\nu_{A,\parallel}^{(\rm{mf})}}$ and $R_{\perp}\sim N^{\nu_{A,\perp}^{(\rm{mf})}}$ with the well known mean-field animal exponents
\begin{equation}
\nu_{A,\parallel}^{(\rm{mf})}=\frac{1}{2}\,,\qquad\nu_{A,\perp}^{(\rm{mf})}=\frac{1}{4}\,.
\end{equation}

\subsection{The stochastic functional for directed branched polymers}

The response functional $\mathcal{J}_{z}=\mathcal{J}_{\rm{DP}}+z\mathcal{M}$, Eq.\ (\ref{Jz}), is form-invariant under three continuous transformations. Thus, three parameters of the functional are redundant. Here, we exploit the symmetries to eliminate two redundant parameters and to define scale-invariant effective couplings.

The symmetry transformations are: (i) a rescaling of the fields
\begin{subequations}
\label{Form-Inv}
\begin{equation}
\tilde{n}\rightarrow\alpha^{-1}\tilde{n}\,,\qquad n\rightarrow \alpha n\,, \label{Resk}
\end{equation}
(ii) a mixing of the fields
\begin{equation}
\tilde{n}\rightarrow\tilde{n}+\beta n\,,\qquad n\rightarrow n\, , \label{Mix}
\end{equation}
and (iii) a shift of the response field
\begin{equation}
\tilde{n}\rightarrow\tilde{n}+\gamma\,,\qquad n\rightarrow n\,. \label{Shift}
\end{equation}
\end{subequations}
We eliminate the redundant parameters $r$ and $c$ by specializing the latter two, which do not transform the density field $n$, by
\begin{equation}
\tilde{n}\rightarrow\tilde{n}+\frac{(r-\tau)}{g}\bigl(1-cn\bigr)\,. \label{Red-Entf}
\end{equation}
This special shift reminds of the saddle-point, Eq.\ (\ref{PhiMF}), and eliminates the uncritical parameter $r$ in favor of  $\tau$. After this shift the fields and their expectation values are small quantities for small $\tau$, and a perturbation expansion is appropriate. The parameter $\tau$ is a free parameter at this place, and, of course, we can set this parameter to zero. However, it is more economical to use it in the renormalization program as a small control parameter because it attains at least an additive renormalization. The shift alone introduces besides the bilinear gradient term $\sim \tilde{n}\nabla^2 n$ a further quadratic gradient term $\sim n\nabla^2 n$ that is thereafter eliminated by the special mixing in Eq.\ (\ref{Red-Entf}). This term also attains renormalizations, and to cure UV-divergencies a counter term $\sim n\nabla^2 n$ must be reintroduced via the renormalization scheme, which we will do as we proceed.

The transformation (\ref{Red-Entf}) suggests the definitions of new coupling constants
\begin{align}
g_0&=g\,,\qquad g_1=g^{\prime}+2cr\,,\nonumber\\
g_2&=r\bigl(f^{\prime}-3cg^{\prime}-3c^2r)/g\,, \label{neueKoppl}
\end{align}
and new control parameters
\begin{align}
\hat\tau_0&=\tau\,,\qquad \hat\tau_1=(r-\tau)(g^{\prime}-2c\tau)/g\,,\nonumber\\
\hat{h}&=z+\bigl(r^2-\tau^2\bigr)/2g\,. \label{neueKontr}
\end{align}
Note that the three control parameters $\hat\tau_0$, $\hat\tau_1$, and $\hat{h}$ go to zero in the mean-field theory of the collapse of large branched polymers, i.e., near $z_c$, and that, however, the coupling constant $g_1$ stays finite.

The elimination of all terms that are at least irrelevant (in the sense of a naive scaling consideration) in comparison to the retained ones reduces the stochastic functional to
\begin{align}
\mathcal{J}^{\prime}_z  &  =\int d^{d}xdt\,\lambda\Big\{\tilde{n}
\Big[\lambda^{-1}\partial_{t}+(\hat\tau_0-\nabla^{2})\Big]n+\frac{\hat\tau_1}{2}n^2\nonumber\\
&\qquad+\Big[-\frac{g_0}{2}\tilde{n}^{2}n+\frac{g_1}{2}\tilde{n}n^{2}
+\frac{g_2}{6}n^{3}\Big] +\hat{h}n\Big\}\,. \label{Jz-neu}
\end{align}
At this place we remark that for $g_2=0$ the stochastic functional $\mathcal{J}_z^{\prime}$ has the same form as the stochastic functional for ordinary directed percolation with a quadratic rapidity-reversal symmetry breaking $\sim \hat\tau_1$. Thus, in this special case the collapse transition belongs to the DP-universality class. Hence, we want to show wether $g_2$ is a relevant perturbation or it goes to zero under renormalization.

The modifications of the dynamical functional including the elimination of the irrelevant higher order couplings make it worthwhile to discuss the support of the functional integration over the fields $\tilde{n}$ and $n$ anew. Like before, the integration of
$n$, the density of monomers, goes along the positive real axes. We write the third order terms of the
integrand of $\mathcal{J}_z$ as
\begin{align}
&\Big[-\frac{g_{0}}{2}\tilde{n}^2 n+\frac{g_{1}}{2}\tilde{n}n^2
+\frac{g_{2}}{6}n^{3}\Big]
\nonumber\\
&=\frac{n}{2}\Big[-g_{0}\Big(\tilde{n} -\frac{g_{1}}{2g_{0}}n\Big)^{2}
+\Big(\frac{g_{2}}{3}+\frac{g_{1}^{2}}{4g_{0}}\Big)n^{2}\Big]\,.
\end{align}
This form shows that the support of integration of $\tilde{n}$ must be so that
$\tilde{n}-g_{1}n/2g_{0}$ is imaginary if this variable becomes large.
The convergence of the functional integral then requires
\begin{equation}
g_{0}\geq0\,,\qquad4g_{0}^{3}g_{2}+3(g_{0}g_{1})^{2}\geq0\, \label{IntegrBdg}
\end{equation}
to make sense beyond perturbation theory. If these conditions are violated, higher order contributions become relevant for stability reasons, and a first-order transition takes place. Next, we consider the quadratic terms of $\mathcal{J}^{\prime}_z$. Writing these terms as
\begin{align}
\tilde{n}\big[\hat\tau_0-\nabla^{2}\big]n&+\frac{\hat\tau_1}{2}n^2
=\Big(\tilde{n} -\frac{g_{1}}{2g_{0}}n\Big)\big[\hat\tau_0-\nabla^{2}\big]n\nonumber\\
&+\frac{1}{2}n\Big[\Big(\hat\tau_1+\frac{g_{1}}{g_{0}}\hat\tau_0\Big)-\frac{g_{1}}{g_{0}}\nabla^{2}\Big]n
\end{align}
we see that we have to require
\begin{equation}
\hat\tau_0\geq0\,,\qquad g_{0}^{2}\hat\tau_1+(g_{0}g_{1})\hat\tau_0\geq
0\,,\qquad(g_{0}g_{1})\geq0\ \label{StabBdg}
\end{equation}
for stability of the saddle-point at $\tilde{n}=n=0$
(for $\hat{h}=\hat{h}_c=0$). If these conditions are violated, a phase transition takes place.

Through the $\delta$-function condition in Eq.~(\ref{Prob_N}) which selects
from all the generated clusters only the clusters with given mass $N$, the stochastic functional $\mathcal{J}^{\prime}_{z}$  has lost its typical causal structure, whereas the absorbing
state condition still holds, $\mathcal{J}^{\prime}_{z}[\tilde{n},n=0]=0$. To restore the
causal structure, we perform the duality transformation
\begin{align}
\label{dualityTrafo}
n(t)=i\tilde
{\varphi}(-t)\, ,  \qquad \tilde{n}(t)=-i\varphi(-t)\, .
\end{align}
These steps lead us from Eq.~(\ref{Jz-neu}) to
\begin{align}
\mathcal{J}  &  =\int d^{d}xdt\,\lambda\Big\{\tilde
{\varphi}\Big[\lambda^{-1}\partial_{t}+(\hat\tau_0-\nabla^{2})\Big]\varphi
-\frac{\hat\tau_1}{2}\tilde{\varphi}^2\nonumber\\
& \qquad +i\Big[\frac{g_0}{2}\tilde{\varphi}\varphi^{2}+\frac{g_1}{2}\tilde{\varphi}^{2}\varphi
-\frac{g_2}{6}\tilde{\varphi}^{3}\Big]+i\hat{h}\tilde{\varphi}\Big\}\,. \label{J_ende}
\end{align}
Evidently, this new functional obeys causality, $\mathcal{J}[\tilde{\varphi}=0,\varphi]=0$, but the absorbing state property is satisfied only if $h=\tau_1=g_2=0$. If only $g_2=0$ holds, it defines RFT with added linear and quadratic symmetry breaking terms. In the following, two facts will play an important role: (i) RFT is equivalent to ordinary DP, and (ii) the added symmetry breaking terms do not alter the fixed point structure of RFT.

As it stands, the $\mathcal{J}$ is  a truly proper field theoretic stochastic functional in the sense of being a minimal model. Nevertheless, it still contains that much generality that allows to study the statistics of swollen DBP and the CDBP transition. These will be the subjects of the following sections.

\section{A brief view on swollen directed branched
polymers}
\label{sec:nonCollapsing}

In the previous section we have learned {\em inter alia} that the shape of $\mathcal{P} (N)$ for large $N$ is controlled by the immediate vicinity of the branch point $z_c$ and that correlation lengths  are large when the parameter $\tau$ defined in Eq.~(\ref{SPL}) is small. For studying large clusters based on our functional $\mathcal{J}$, Eq.\ (\ref{J_ende}), this means that we should focus on the limit of vanishing control parameters $\hat{h}$ and $\hat\tau_0$. The collapse transition of these large polymers occurs for vanishing $\hat\tau_1$. This transition will be treated in the later sections. Here we are interested in the swollen phase of these polymers, and hence we consider a $\hat\tau_1$ that is positive and finite, say of the order of $1$. Then, $\hat\tau_1$ can be eliminated through a simple redefinition of the fields $(\hat\tau_1/2)^{1/2}\tilde{\varphi}\rightarrow\tilde{\varphi}$, $(\hat\tau_1/2)^{-1/2}\varphi\rightarrow\varphi$, $(\hat\tau_1/2)^{1/2}g_0\rightarrow g_0$, $(\hat\tau_1/2)^{-1/2}g_1\rightarrow g_1$, $(\hat\tau_1/2)^{-3/2}g_2\rightarrow g_2$, or formally by setting $\hat\tau_1=2$.

Now, we count engineering dimensions. As usual, we employ the some inverse length scale $\mu$ which is convenient for this task. Recalling that $d$ denotes the number of transversal dimensions only, we have
\begin{subequations}
\begin{align}
&\varphi \sim\mu^{(d-2)/2}\,,\qquad\tilde{\varphi}\sim\mu^{(d+2)/2}\,,\\
&\hat\tau_0 \sim\mu^{2}\,,\qquad\qquad\; g_0\sim\mu^{(6-d)/2}\,,\\
&g_1 \sim\mu^{(2-d)/2}\,,\qquad g_2\sim\mu^{-(d+2)/2}\,.
\end{align}
\end{subequations}
We see that the upper critical dimension $d_{c}$ is $6$ and that $g:=g_0$ is the only relevant coupling constant below $d_c$. Neglecting the irrelevant couplings, we obtain the response
functional
\begin{align}
\mathcal{J}_{YL}&=\int d^{d}xdt\,\Big\{\tilde{\varphi}\Big[\partial_{t}\varphi
+\lambda\frac{\delta\mathcal{H}_{YL}}{\delta\varphi}\Big]
-\lambda\tilde{\varphi}^{2}\Big\}\,,\label{J_YL}
\end{align}
that is known to describe the relaxational dynamics at the Yang-Lee singularity edge \cite{BrJa81}. Here,
\begin{align}
\mathcal{H}_{YL}  &  =\int d^{d}x\Big\{\frac{1}{2}(\nabla\varphi)^{2}
+\frac{\tau}{2}\varphi^{2}+i\frac{g}{6}\varphi^{3}+i\hat{h}\varphi\Big\}\,.
\label{H_YL}
\end{align}
is the Hamiltonian of an Ising-order parameter in an
imaginary field near criticality \cite{Fi78} with $\tau = \hat\tau_0$. It was shown \cite{BrJa82,Ca82}
that the entropic exponent of the probability $\mathcal{P}(N)\sim N^{-\dot{\theta}
}\exp(\mathrm{-}N/N_{0})$ and the exponent of the transversal gyration radius
$R_{\perp}\sim N^{\dot{\nu}_{A,\perp}}$ are determined (as conjectured by Day and
Lubensky \cite{DaLu82}) by
\begin{align}
\dot{\theta} &  =1+\sigma_{YL}\,,\\
\dot{\nu}_{A,\perp}  &  =\dot{\theta}/d\,,
\end{align}
where $\sigma_{YL}$ is the Yang-Lee exponent which relates the order parameter
and the imaginary magnetic field via $(\Phi-\Phi_{c})\sim\left\vert
\hat{h}-\hat{h}_{c}\right\vert ^{\sigma_{YL}}$. The ($\varepsilon=6-d$)-expansion of the
longitudinal Flory exponent is \cite{BrJa81,BrJa82}
\begin{equation}
\dot{\nu}_{A,\parallel}=\frac{1}{2}+\left[1+\left(  \frac{95}{108}-\frac{9}{2}
\ln\frac{4}{3}\right)\frac{\varepsilon}{6}\right]\frac{\varepsilon}
{24}+O(\varepsilon^{3})\,. \label{BrJa_eps}
\end{equation}
Table~\ref{table1} compiles values for $\dot{\theta}$ and $\dot{\nu}_{A,\parallel}$ based on exact results for $d=0$, $1$ and $d=2$ \cite{Ca85} and the third order $\varepsilon$-expansion results of
de Alcantara Bonfim et al.~\cite{AlKiMc81}. For corrections to scaling and the derivation of the correction exponent in $\varepsilon$-expansion see \cite{Bre84,LaFi95}.
\begin{table}
\begin{tabular}
[c]{l|lllllll}
$d:\quad$ & $\;\;0\qquad$ & $\;1\qquad$ & $\;2\qquad$ & $\;3\qquad$ &
$\;4\qquad$ & $\;5\qquad$ & $\;6\quad$\\\hline\hline
$\dot{\theta}:\quad$ & $\;\;0$ & $1/2$ & $5/6$ & $1.08$ & $1.26$ & $1.40$ & $3/2$\\
$\dot{\nu}_{A,\parallel}:\quad$ & $\;\;1$ & $0.79$ & $0.69$ & $0.63$ & $0.58$ &
$0.54$ & $1/2$
\end{tabular}
\caption{Values for the critical exponents $\dot{\theta}$ and $\dot{\nu}_{A,\parallel}$ for various transversal dimensions $d$.}
\label{table1}
\end{table}

\section{Field theory of collapsing directed branched polymers}
\label{sec:fieldTheory}

Now, we return to the CDBP transition as the main topic of our paper.

\subsection{Renormalization}
\label{subsec:renormalization}

Our response functional $\mathcal{J}$, Eq.~(\ref{J_ende}), describes specifically the CDBP transition when both control parameters $\hat\tau_0$ and $\hat\tau_1$ become critical. In this case, they  scale in terms of the inverse length scale $\mu$ as $\hat\tau_0\sim\hat\tau_1\sim\mu^{2}$. Then the engineering dimensions of the fields and the remaining parameters follow as
\begin{align}
\varphi &  \sim\mu^{d/2}\,,\qquad\tilde{\varphi}\sim\mu^{d/2}\,,\\
\hat\tau_0  &  \sim\hat\tau_1\sim\mu^{2}\,,\qquad \hat{h}\sim\mu^{(4+d)/2}\, ,\\
g_0  &  \sim g_1\sim g_2\sim\mu^{(4-d)/2}\,.
\end{align}
Now, the upper critical dimension is $d_{c}=4$. All three coupling constants are relevant below
$d_{c}$. There are no further contributions that are relevant under the condition of causality
(originating from the absorbing condition of the process) which we have restored through the duality transformation~(\ref{dualityTrafo}).

It is well known that a perturbation expansion of correlation and response
functions (generally called Greens functions) based on a field theoretic model
like $\mathcal{J}$ produces UV-singularities which have to
be regularized and renormalized by singular counter terms. Here, we use minimal renormalization, i.e., dimensional regularization followed by minimal subtraction of $\varepsilon$-poles, to cure the theory from UV-divergencies. For the general principles and methods of renormalization theory see, e.g., Ref.~\cite{Am84,ZJ02}.

In the following, we use a ring $\mathring{}$ to mark bare (unrenormalized) quantities, i.e., we let  $\varphi\rightarrow\mathring{\varphi}$, and so on. From here on, quantities without a ring are understood as renormalized quantities. To get rid of $\varepsilon$-poles, we use the renormalization scheme
\begin{subequations}
\label{Ren}
\begin{align}
& \mathring{\varphi}   =Z^{1/2}(\varphi+K\,\tilde{\varphi}),\qquad
\mathring{\tilde{\varphi}}=Z^{1/2}\tilde{\varphi}\,,\label{Ren-1}\\
& \mathring{\lambda}   =Z^{-1}Z_{\lambda}\lambda\,,\qquad
\underline{\mathring{\hat\tau}}=Z_{\lambda}^{-1}\underline{\underline{Z}}\cdot\underline{\hat\tau}
+\underline{\mathring{\hat\tau}}_c \,,\label{Ren-2}\\
& \mathring{\hat{h}}  = Z^{1/2}Z_{\lambda}^{-1}\big(\hat{h}+\frac{1}{2}G_{\varepsilon}^{1/2}
\mu^{-\varepsilon/2}\underline{\hat\tau}\cdot\underline{\underline{A}}
\cdot\underline{\hat\tau}\big)
+\mathring{\hat{h}}_c+\underline{\mathring{C}}\cdot\underline{\hat\tau}\,,\label{Ren-3}\\
 &\mathring{g}_{\alpha}  = Z^{-1/2}Z_{\lambda}^{-1}G_{\varepsilon}^{-1/2}\mu
^{\varepsilon/2}\left(  u_{\alpha}+B_{\alpha}\right)  \,, \label{Ren-4}
\end{align}
\end{subequations}
 where $\varepsilon=4-d$ and $G_{\varepsilon}=\Gamma(1+\varepsilon/2)/(4\pi)^{d/2}$. Here, we have introduced the two-dimensional vector $\underline{\hat\tau}=(\hat\tau_0,\hat\tau_1)$. Note that the mixing-term proportional to $K$ reintroduces the aforementioned gradient term, see the discussion below Eq.~(\ref{Red-Entf}). In minimal renormalization with dimensional expansion   the additive contributions $\underline{\mathring{\hat\tau}}_c$, $\mathring{\hat{h}}_c$, $\underline{\mathring{C}}$ become formally zero, and the  $\varepsilon$-content of the other counterterms are defined by pure Laurent-series:

\begin{equation}
Z_{..}-1=:Y_{..}=\sum_{k=1}^{\infty}\frac{Y_{..}^{(k)}}{\varepsilon^{k}}
=\sum_{k=1}^{\infty}\varepsilon^{-k}\sum_{l=k}^{\infty}Y_{..}^{(l,k)}\,,
\label{minRen}
\end{equation}
where the counterterms $Y_{..}^{(l,k)}$, when determined at the loop-order $l$, are
homogeneous polynomials of the renormalized dimensionless coupling constants
$u_{\alpha}$ of order $2l$: $Y_{..}^{(l,k)}(su_{\alpha})=s^{2l}Y_{..}
^{(l,k)}(u_{\alpha})$. Corresponding expansions are valid for the various
other counterterms $K$, $\underline{\underline{A}}$, $B_{\alpha}$ which themselves
are homogeneous polynomials of order $2l$, $2l-1$, $2l+1$, respectively, at
each loop-order $l$. We calculate the various renormalization factors $Z_{..}$ and additional counter-terms using a two-tiered approach, viz.\ we carry out an explicit 1-loop calculation and we employ Ward identities. The latter will be discussed further below. Our 1-loop calculation, see the Appendix for details, produces
\begin{subequations}
\label{Ren-1L}
\begin{align}
Z  &  =1+\frac{u_{0}u_{1}}{4\varepsilon}+\ldots\,,\quad
Z_{\lambda}=1+\frac{u_{0}u_{1}}{8\varepsilon}+\ldots\,,\label{Ren-1L-2}\\
\underline{\underline{Z}}  &  =\underline{\underline{1}}+\frac{u_{0}}{4\varepsilon}
\begin{pmatrix}
2u_{1} & -2u_{0}\\
3u_{2} & 4u_{1}
\end{pmatrix}
+\ldots\,,\\
K&=\Big(-\frac{u_{0}}{8\varepsilon}+\ldots\Big)u_{2}\,,\quad\underline{\underline{A}}
=\frac{u_{0}}{2\varepsilon}
\begin{pmatrix}
0 & 1\\
1 & 0
\end{pmatrix}
+\ldots\,,\label{Ren-1L-4}\\
B_{0}  &  =\Big(\frac{u_{0}u_{1}}{\varepsilon}+\ldots\Big)u_{0}\,,\quad B_{2}=\Big(\frac{21u_{1}}{8\varepsilon}+\ldots\Big)u_{0}u_{2}\,,\\
B_{1} &=\Big(\frac{4u_{1}^{2}-3u_{2}u_0}{4\varepsilon}+\ldots\Big)u_{0}
\,,\label{Ren-1L-8}
\end{align}
\end{subequations}
where ellipsis symbolize terms of higher loop-order which are regular expansions in the coupling constants $u_{\alpha}$ but, of course singular in $\varepsilon$. Note that in the case $u_2=0$, where rapidity inversion symmetry holds, $K$ and $B_2$ are zero to all orders. In this case we have also $B_0/u_0=B_1/u_1$. In the general case all renormalizations are trivial if $u_0=0$ because then only diagrams without loops are generated by the perturbation expansion.

\subsection{The renormalization group}

Next, we derive a renormalization group (RG) in the usual way by
utilizing that the bare theory is independent of the external, arbitrary
inverse length scale $\mu$ which enters through the renormalization scheme. Because of this independence, the derivative $\mu\partial/\partial\mu\mathcal{\mathring{Q}}|_{0}$ ($|_{0}$ denotes
derivatives holding bare parameters fixed) vanishes for each unrenormalized quantity $\mathcal{\mathring{Q}}$. Switching from bare to renormalized quantities $\mathcal{Q}$ depending only on renormalized parameters, the derivative $\mu\partial/\partial\mu|_{0}$ changes to
the RG differential operator
\begin{align}
\mathcal{D}_{\mu} &=\mu\partial_{\mu}+\zeta\lambda\partial_{\lambda
}+\underline{\hat\tau}\cdot\underline{\underline{k}}\cdot\partial_{\underline
{\hat\tau}}+{\textstyle\sum\nolimits_{\alpha}}\beta_{\alpha}\partial_{u_{\alpha}}\nonumber\\
& +\left[  (\gamma/2-\zeta)\hat{h}+G_{\varepsilon}^{1/2}\mu^{-\varepsilon
/2}(\underline{\hat\tau}\cdot\underline{\underline{a}}\cdot\underline{\hat\tau
})/2\right]  \partial_{\hat{h}}\,, \label{RGE-Diff}
\end{align}
which defines the infinitesimal generator of the RG, the (half-) group of scale changes of the length $\mu^{-1}$. The application of $\mathcal{D}_{\mu}$ to renormalized fields in averages produces
\begin{equation}
\mathcal{D}_{\mu}\tilde{\varphi}=-\frac{\gamma}{2}\tilde{\varphi}\,,\qquad
\mathcal{D}_{\mu}\varphi=-\frac{\gamma}{2}\varphi-\frac{\gamma^{\prime}}{2}\tilde{\varphi}\,. \label{RG-Feld}
\end{equation}
Recalling the renormalization scheme~(\ref{Ren}) and the form of the renormalization factors $Z_{..}$ and the additional counter-terms discussed below it, we find
\begin{subequations}
\label{RGEfunctions}
\begin{align}
\gamma_{..}&=\left.  \mu\partial_{\mu}\ln Z_{..}\right\vert _{0}=-\frac{1}{2}(u\cdot\partial_u)
Y_{..}^{(1)}=-\sum_{l=1}^{\infty}lY_{..}^{(l,1)}\,,\label{Wils-Fu}\\
\gamma^{\prime}&=-(u\cdot\partial_u) K^{(1)}=-2\sum_{l=1}^{\infty}lK^{(l,1)}\,,\\
\underline{\underline{a}}&=\frac{1}{2}\left((u\cdot\partial_u)+1\right)\underline{\underline{A}}^{(1)}=\sum_{l=1}^{\infty
}l\underline{\underline{A}}^{(l,1)}\,,\label{alph-Fu}\\
b_{\beta}&=\frac{1}{2}\left((u\cdot\partial_u)-1\right)
B_{\alpha}^{(1)}=\sum_{l=1}^{\infty}lB_{\alpha}^{(l,1)}\,,
\end{align}
\end{subequations}
where $(u\cdot\partial_u)={\textstyle\sum\nolimits_{\alpha}} u_{\alpha}\partial_{u_{\alpha}}$, as well as
\begin{subequations}
\begin{align}
\zeta &  =\left.  \mu\partial_{\mu}\ln\lambda\right\vert _{0}=\gamma
-\gamma_{\lambda}\,,\label{zet-lamb}\\
\underline{\hat\tau}\cdot\underline{\underline{k}}  &  =\left.  \mu\partial_{\mu
}\underline{\hat\tau}\right\vert _{0}=(\gamma_{\lambda}-\underline{\underline
{\gamma}})\cdot\underline{\hat\tau}\,,\label{kappa-Fu}\\
\beta_{\alpha}  &  =\left.  \mu\partial_{\mu}u_{\alpha}\right\vert
_{0}=\left(  -\frac{\varepsilon}{2}+\frac{\gamma}{2}+\gamma_{\lambda}\right)
u_{\alpha}+b_{\alpha}\,,\label{beta-Fu}\\
\left.  \mu\partial_{\mu}\hat{h}\right\vert _{0}  &  =(\gamma_{\lambda}
-\gamma/2)\hat{h}+G_{\varepsilon}^{1/2}\mu^{-\varepsilon/2}(\underline{\hat\tau}
\cdot\underline{\underline{a}}\cdot\underline{\hat\tau})/2\,. \label{h-Fu}
\end{align}
\end{subequations}
for the RG functions. Explicit 1-loop results for these RG functions follow readily  from Eqs.~(\ref{Ren-1L}).

\subsection{Ward identities, redundancy, and the invariant renormalization group equations}
\label{subsec:Wardetc}
It has been noted since the early days of field theory that each symmetry transformation (form-invariance) which transforms not only fields but also parameters of a field theoretic functional implies the redundancy of one of its parameters and vice versa~\cite{wegner76}. Redundancy means that a respective parameter is related to other parameters of the functional through simple linear transformations of the fields. These transformations, however, must not change the physical contents of the functional. Hence, in a proper field theory, redundant parameters can be and must be eliminated through linear transformations of the fields.

As we have already noted, the response functional $\mathcal{J}$, Eq.\ (\ref{J_ende}), is form-invariant under the three continuous transformations (\ref{Form-Inv}). The mixing and shift transformations, Eq.\ (\ref{Red-Entf}), were used to eliminate two of the corresponding redundant parameters, $r$ and $c$, and they introduced $\tau=\hat\tau_0$ as a small free parameter. Now, we use this freedom to derive Ward identities from the shift transformation, Eq.\ (\ref{Shift}), which reads
\begin{equation}
\varphi\rightarrow\varphi^{\prime}=\varphi+i\gamma\,,
\qquad{\tilde{\varphi}}\rightarrow{\tilde{\varphi}}^{\prime}={\tilde{\varphi}}\,, \label{Shift1}
\end{equation}
when expressed in terms of the fields $\varphi$ and $\tilde{\varphi}$.
The response functional $\mathcal{J}$, Eq.~(\ref{J_ende}), is invariant under this shift,
\begin{equation}
\mathcal{J}[{\tilde{\varphi}},\varphi;\underline{\hat\tau},g_{\alpha},h]
=\mathcal{J}[{\tilde{\varphi}}^{\prime},\varphi^{\prime};\underline{\hat\tau}^{\prime},g_{\alpha},
h^{\prime}]\,, \label{Shift-Inv-Fu}
\end{equation}
if we also change the control parameters as follows:
\begin{subequations}
\label{shift-param}
\begin{align}
\underline{\hat\tau}  &  \rightarrow\underline{\hat\tau}^{\prime}=\underline{\hat\tau
}+\gamma\underline{{\bar{g}}}\,,\label{shift-tau}\\
\hat{h}  &  \rightarrow  \hat{h}^{\prime}= \hat{h}-\gamma{\hat\tau}_0-\frac{\gamma^{2}}{2} g_{0}\,.
\label{shift-h}%
\end{align}
\end{subequations}
Here, we have introduced the two-dimensional vector $\underline{\bar{g}}=(g_{0},-g_{1})$. Primarily, the shift is defined for unrenormalized quantities. However, because it does not involve a transformation of the coupling constants, and the renormalization constants are only functions of the couplings, the shift-invariance, Eq.\ (\ref{Shift-Inv-Fu}), also holds for the renormalized quantities with $g_{\alpha}:= G_{\varepsilon}^{-1/2}u_{\alpha}\mu^{\varepsilon/2}$.

Now, we compare the bare transformations with their renormalized counterpart. We renormalize the bare form of Eq.~(\ref{shift-tau}),
${\underline{\mathring{\hat\tau}}}^{\prime}-\underline{\mathring{\hat\tau}}
=\mathring{\gamma}\underline{\mathring{g}}$ with $\mathring{\gamma}=Z^{1/2}\gamma$:
\begin{equation}
\underline{\underline{Z}}\cdot\left(\underline{\hat\tau}^{\prime}-\underline
{\hat\tau}\right)  =\gamma G_{\varepsilon}^{-1/2}\mu^{\varepsilon/2}\left(
\underline{\bar{u}}+\underline{\bar{B}}\right)  \,,
\end{equation}
where $\underline{\bar{u}}=(u_{0,}-u_{1})$ and $\underline{\bar{B}}
=(B_{0,}-B_{1})$. Splitting this equation in singular and non-singular parts,
we get
\begin{subequations}
\begin{align}
\underline{\hat\tau}^{\prime}  &  =\underline{\hat\tau}+\gamma G_{\varepsilon}
^{-1/2}\mu^{\varepsilon/2}\underline{\bar{u}}\,,\label{Ren-shift-1}\\
\underline{\bar{B}}  &  =\underline{\underline{Y}}\cdot\underline{\bar{u}}\,,
\label{Ward-1}
\end{align}
\end{subequations}
with $\underline{\underline{Y}}=\underline{\underline{Z}}-\underline{\underline{1}}$. The
last equation, the first Ward-identity, is easily checked at $1$-loop order by using
Eqs~(\ref{Ren-1L}). We perform the same procedure with Eq.~(\ref{shift-h})
and obtain
\begin{subequations}
\begin{align}
\hat{h}^{\prime}  &  =\hat{h}-\gamma\hat\tau_0-\frac{\gamma^{2}}{2}G_{\varepsilon}^{-1/2}
\mu^{\varepsilon/2}u_{0}\,,\label{Ren-shift-2}\\
\underline{Y}  &  =-\underline{\bar{u}}\cdot\underline{\underline{A}}\,,
\label{Ward-2}
\end{align}
\end{subequations}
with $\underline{Y}=(Y_{0,0},Y_{0,1})$. Also the second Ward-identity,
Eq.~(\ref{Ward-2}), is easily checked at $1$-loop order by using
Eqs~(\ref{Ren-1L}).

The Ward identities (\ref{Ward-1}) and (\ref{Ward-2}) imply that there are further relations between the RG functions, viz.\
\begin{equation}
\underline{\beta}=\frac{1}{2}(-\varepsilon+\gamma)\underline{u}+\underline
{\bar{u}}\cdot\underline{\underline{k}}\,,\quad\underline{\gamma}
=\underline{\bar{u}}\cdot\underline{\underline{a}}\,, \label{Ward-ren}
\end{equation}
where $\underline{\gamma}=(\gamma_{00},\gamma_{01})$. These relations can be used {\em inter alia} to check the consistency of the 1-loop results for the various RG functions and the results~(\ref{Ren-1L}) leading to them. Our results fulfill this consistency check. More generally we derive relations between different vertex functions. Note that the vertex function $-\Gamma_{\tilde{k},k}$ is diagrammatically defined as the sum of all one-line irreducible diagrams with $\tilde{k}$ amputated external $\tilde{\varphi}$- and ${k}$
amputated external ${\varphi}$-lines \cite{ZJ02}. Let $\Gamma[\tilde{\varphi},\varphi;\underline{\hat\tau},g_{\alpha},\hat{h}]$ be the generating functional of the vertex functions, which we call the dynamic free-energy functional. It has the same invariance property, Eq.\ (\ref{Shift-Inv-Fu}), as the stochastic functional $\mathcal{J}$, which is identical to the mf-approximation of $\Gamma$. Hence, we have
\begin{align}
\Gamma[&{\tilde{\varphi}},\varphi;\underline{\hat\tau},g_{\alpha},\hat{h}]
=\Gamma[{\tilde{\varphi}},\varphi;\underline{\hat\tau},g_{\alpha},0]+\lambda \hat{h}\int d^{d}xdt\tilde\varphi({\bf{x}},t)\nonumber\\
&=\Gamma[{\tilde{\varphi}},\varphi+i\gamma;\underline{\hat\tau}+\gamma \underline{\bar{g}},g_{\alpha},
\hat{h}-\gamma\hat\tau_0-\gamma^2g_0/2]\,, \label{Shift-Gamma-Fu}
\end{align}
Setting $\gamma=0$ after differentiation leads to
\begin{equation}
\int d^{d}xdt\frac{\delta\Gamma}{\delta\varphi({\bf{x}},t)}
=i\Big(g_0\frac{\partial}{\partial\hat\tau_0}-g_1\frac{\partial}{\partial\hat\tau_1}
-\hat\tau_0\frac{\partial}{\partial\hat{h}}\Big)\Gamma\,, \label{Ward-Gamma}
\end{equation}
which is sometimes called equation of motion. This general property of the dynamic free-energy energy functional holds equally well in renormalized and unrenormalized form, and is very helpful for higher-loop calculations. Especially, we find by taking repeated functional derivatives
\begin{subequations}
\label{Ward-Gamma-spec}
\begin{align}
\Gamma_{1,1}({\bf{0}})&=\hat\tau_0 + i\Big(g_0\frac{\partial}{\partial\hat\tau_0}
-g_1\frac{\partial}{\partial\hat\tau_1}\Big)\Gamma_{1,0}({\bf{0}})\,,\label{Ward-Gamma-spec-1}\\
\Gamma_{1,2}({\bf{0}})&= i\Big(g_0\frac{\partial}{\partial\hat\tau_0}
-g_1\frac{\partial}{\partial\hat\tau_1}\Big)\Gamma_{1,1}({\bf{0}})\,,\label{Ward-Gamma-spec-2}\\
\Gamma_{2,1}({\bf{0}})&= i\Big(g_0\frac{\partial}{\partial\hat\tau_0}
-g_1\frac{\partial}{\partial\hat\tau_1}\Big)\Gamma_{2,0}({\bf{0}})\,,\label{Ward-Gamma-spec-}
\end{align}
\end{subequations}
where the argument ${\bf 0}$ means that the wavevectors and frequencies of the vertex functions are set to zero.

Up to this point we have not used the scaling transformation, Eq.\ (\ref{Resk}), for eliminating a further redundancy. Applying this rescaling of the fields, $\mathcal{J}$ remains invariant if we change its parameters to
\begin{align}
\hat\tau_0  &  \rightarrow\hat\tau_0\,,\qquad\hat\tau_1\rightarrow\alpha^{2}\hat\tau
_{1}\,,\qquad \hat{h}\rightarrow\alpha \hat{h}\,,\nonumber\\
g_{0}  &  \rightarrow\alpha^{-1}g_{0}\,,\qquad g_{1}\rightarrow\alpha
g_{1}\,,\qquad g_{2}\rightarrow\alpha^{3}g_{2}\,. \label{Resk1}
\end{align}
This invariance holds likewise for the renormalized and unrenormalized theory. Hence,
it would be absolutely wrong to search at this place for fixed points, i.e., zeros of of all  three Gell-Mann--Low functions, $\beta_{\alpha}=0$, because the free scale transformation poisons the renormalization flow. Before we can search for fixed points, we first must eliminate the scaling redundancy through the definition of scale invariant quantities. We do that by defining  the invariant dimensionless coupling constants
\begin{subequations}
\label{inv-Quant}
\begin{equation}
u=u_{0}u_{1}\,,\quad w=u_{0}^{3}u_{2}\,, \label{inv-Coupl1}
\end{equation}
and changing the fields and control parameters to their scaling invariant counterparts
\begin{align}
&\tilde\phi = g_{0}^{-1}\tilde{\varphi} \,,\quad \phi=g_{0}\varphi\,,
\label{inv-Par-1}\\
&\tau_{0}  =\hat\tau_0\,,\quad \tau_{1}= u_{0}^{2}\hat\tau_1\,,\quad
h  = g_{0}\hat{h}\,. \label{inv-Par-2}
\end{align}
\end{subequations}

The RG differential operator $\mathcal{D}_{\mu}$ applied only to scaling invariant quantities reduces to
\begin{align}
\mathcal{D}_{\mu}&=\mu\partial_{\mu}+\zeta\lambda\partial_{\lambda
}+\underline{\tau}\cdot\underline{\underline{\kappa}}\cdot\partial
_{\underline{\tau}}+\beta_{u}\partial_{u}+\beta_{w}\partial_{w}\nonumber\\
&  +\left[  (\gamma_{1}/2-\zeta)h+(\underline{\tau}\cdot\underline{\underline{\alpha}}
\cdot\underline{\tau})\right]  \partial_{h}\,, \label{inv-RGE-Diff}
\end{align}
and its application to the invariant fields in averages produces
\begin{equation}
\mathcal{D}_{\mu}\phi=-\frac{\gamma_0}{2}\phi-\frac{\gamma_2}{2}\tilde{\phi}\,,\qquad
\mathcal{D}_{\mu}\tilde{\phi}=-\frac{\gamma_1}{2}\tilde{\phi}\,. \label{inv-RG-Feld}
\end{equation}
Here, we have defined the invariant RG functions
\begin{equation}
\gamma_{0} =\gamma-\varepsilon-\zeta_{0}\,,\quad\gamma_{1} =\gamma+\varepsilon+\zeta_{0}\,,\quad
\gamma_{2} =u_{0}^{2}\gamma^{\prime}\label{inv-gamma}\,,
\end{equation}
where
\begin{equation}
\zeta_{0}=\frac{2\beta_{0}}{u_{0}}\,.\label{zeta_0}
\end{equation}
The invariant matrices $\underline{\underline{\kappa}}$ and $\underline
{\underline{\alpha}}$ follow from their respective counterparts $\underline
{\underline{k}}$ and $\underline{\underline{a}}$ as
\begin{subequations}
\begin{align}
\kappa_{00}  &  =k_{00}\,,\quad\kappa_{10}=u_{0}^{-2}k_{10}\,,\nonumber\\
\kappa_{01} & =u_{0}^{2}k_{10}\,,\quad\kappa_{11}=k_{11}+\zeta_{0}\,,\label{kappa}\\
\alpha_{00}  &  =u_{0}a_{00}\,,\quad\alpha_{10}=\alpha_{01}=u_{0}^{-1}a_{10}\,,\nonumber\\
\alpha_{11}&=u_{0}^{-3}a_{10}\,.\label{alph}
\end{align}
\end{subequations}

From the above, we can straightforwardly collect the following explicit 1-loop for the invariant RG functions:
\begin{subequations}
\label{inv-Wils-eps}
\begin{align}
\gamma_{0}  &  =-\frac{7u}{4}+\ldots\,,\quad\gamma_{1}=\frac{5u}{4}+\ldots\,,\\
\gamma_{2}&=(\frac{1}{4}+\ldots)w\,, \quad \zeta  =-\frac{u}{8}+\ldots\,,\\
\zeta_{0}&=-\varepsilon+\frac{3u}
{2}+\ldots\,,
\end{align}
\end{subequations}
and
\begin{subequations}
\begin{align}
\underline{\underline{\kappa}}  &  =
\begin{pmatrix}
3u/8& 3w/4\\
-1/2 & 19u/8-\varepsilon
\end{pmatrix}
+\ldots\,,\label{inv-kappa}\\
\underline{\underline{\alpha}}  &  =
\begin{pmatrix}
0 & 1/2\\
1/2 & 0
\end{pmatrix}
+\ldots\,. \label{inv-alpha}
\end{align}
\end{subequations}
In the DP case, $w=\tau_1=0$, the RGE has to have the usual form known from DP \cite{Ja01,JaTa05}. It follows that $\kappa_{01}$ and $\alpha_{00}$ have to be zero to all loop orders for $w=0$.
For the invariant Gell-Mann--Low functions we find
\begin{subequations}
\label{inv-beta-Fu}
\begin{align}
\beta_{u}&=\left(-\varepsilon+\frac{3}{2}u\right)u-\frac{3}{4}w+\ldots\,,\label{inv-beta-Fu-u}\\
\beta_{w}&=\left(-2\varepsilon+\frac{37}{8}u+\ldots\right)w\,.
\label{inv-beta-Fu-v}
\end{align}
\end{subequations}

Once again, the Ward identities lead to relations between the RG functions, namely
\begin{equation}
\kappa_{00}=(\gamma-\zeta)-(\underline{\bar{v}}\cdot\underline{\underline{\alpha}})_0\,,\quad
\kappa_{10}=-(\underline{\bar{v}}\cdot\underline{\underline{\alpha}})_1\,, \label{inv-Ward-1}
\end{equation}
and
\begin{equation}
\beta_{u}/u  =-\underline{\bar{v}}\cdot\underline{\underline{\kappa}}\cdot\underline{v}\,,\quad
(\underline{\bar{v}}\cdot\underline{\underline{\kappa}})_{0}  =-\gamma_{0}/2\,, \label{inv-Ward-2}
\end{equation}
where we have introduced the two orthogonal vectors $\underline{v}=(1,u^{-1})$ and
$\underline{\bar{v}}=(1,-u)$. We define
\begin{subequations}
\label{kappa-EV}
\begin{align}
\kappa_{0}  &  =(\underline{\bar{v}}\cdot\underline{\underline{\kappa}}
)_{0}=\kappa_{00}-u\kappa_{10}=\kappa_{11}-u^{-1}\kappa_{01}-\beta
_{u}/u\,,\label{kappa-EV1}\\
\kappa_{1}  &  =(\underline{\underline{\kappa}}\cdot\underline{v})_{0}
=\kappa_{00}+\kappa_{01}u^{-1}=\kappa_{11}+\kappa_{10}u-\beta_{u}/u\,.
\label{kappa-EV2}
\end{align}
\end{subequations}
Hence, if  $\beta_{u}=0$, especially at a fixed point, it follows from
Eq.~(\ref{inv-Ward-2}) that $\underline{\bar{v}}$ and $\underline
{v}$ are a left and a right eigenvector of $\underline{\underline
{\kappa}}$, respectively, since the two vectors are orthogonal.
In this case, we find the two eigenvalues $\kappa_{0}=-\gamma_{0}/2$ and
$\kappa_{1}$. Of course, our 1-loop results for the invariant RG functions satisfy these relations holding to all loop-orders.

To determine the critical behavior, we need to extract those quantities that are invariant under all the symmetry transformations of the theory. Applying the shift transformation the dynamic free energy functional~(\ref{Shift-Gamma-Fu}), that such a set of quantities is given by $\tilde\phi$ and the combinations
\begin{subequations}
\label{vollInv}
\begin{align}
&S=\phi+i\sigma\,,\\
&H=2h-2\tau\sigma-\sigma^2\,,\\
&y=(\underline\tau\cdot\underline{v})=\tau+\sigma\,,
\end{align}
\end{subequations}
where $\sigma=\tau_1/u=g_0\hat\tau_1/g_1$. Note that $H$ is linearly related to the Laplace-variable $z$.

In the following we are mainly interested in the average $\langle S\rangle=i(g_0\Phi+\sigma)=:-iM$ since this quantity is linearly related to $\Phi=\langle\tilde{n}\rangle_{\rm{DP}}$. We recall that the critical part of $\Phi$ yields the asymptotic behavior of the probability $\mathcal{P}(N)$ by an inverse Laplace transformation. Applying Eq.\ (\ref{inv-RG-Feld}) together with $\langle\tilde{\phi}\rangle=0$, and using the relations (\ref{inv-Ward-1}) to (\ref{kappa-EV}) between the RG-functions, we obtain the RGEs
\begin{subequations}
\label{flow}
\begin{align}
\mathcal{D}_{\mu}y  &  =\kappa_{1}y\,,\\
\mathcal{D}_{\mu}M  &  =\kappa_{0}M-u^{-1}\kappa_{01}y\,,\\
\mathcal{D}_{\mu}H  &  =(\kappa_{0}+\gamma_{\lambda})H  +(\alpha_{00}-2u^{-1}\kappa_{01})y^{2}\,.
\end{align}
\end{subequations}

\subsection{Fixed points and critical exponents}
\label{subsec:exponents}

Next, we search for the stable fixed points of the
invariant RGE~(\ref{inv-RG-Feld}) with the RG generator Eq.\ (\ref{inv-RGE-Diff}). The fixed point equations $\beta_{u\ast}=\beta_{w\ast}=0$ possess the following solutions: (i) the trivial fixed point $u_{\ast}=w_{\ast}=0$ which is unstable below four transversal dimensions,
(ii) a fully stable fixed point
\begin{equation}
u_{\ast}=\frac{2\varepsilon}{3}+\ldots\,,\quad w_{\ast}=0\,, \label{stabFP}
\end{equation}
with stability (Wegner-)exponents $\lambda_u = \varepsilon+\ldots$ and $\lambda_w = 13\varepsilon/12+\ldots$  (which correspond to the \emph{two} leading correction-to-scaling exponents), and (iii) a fixed point with one stable and one unstable direction
\begin{equation}
u_{\ast}=\frac{16\varepsilon}{37}+\ldots\,,\quad w_{\ast}=-\frac{13}{3}\left(
\frac{8\varepsilon}{37}\right)  ^{2}+\ldots=-\frac{13}{12}u_{\ast}^{2}%
+\ldots\,, \label{instFP}
\end{equation}
which lies on a separatrix between the region of attraction of the fully
stable fixed point and a region where the renormalization flow tends to
infinity, signalling possibly a discontinuous collapse transition. However,
this fixed point lies in the unstable region, and is thus not accessible by our
model. We conclude, therefore, that the renormalization group flow reaches
asymptotically the fully stable fixed point. The line $w=0$ in the
two-dimensional space of the invariant coupling constants has to be a fixed line of
the renormalization flow because this line corresponds to the RFT-model, Eq.~(\ref{J_ende}) with $g_2=0$,
with its rapidity reversal invariance. Thus, $w_{\ast}=0$ holds to all
loop-orders. We conclude that as long as the collapse transition is continuous,
it belongs to the universality class of directed percolation in all dimensions.

We would like to underscore that this result, which is the main result of our paper, holds to arbitrary order in perturbation theory. The pivotal point is that for $w\to 0$, which is the case at the stable fixed point, $\mathcal{J}$ reduces to RFT and thus, the universal behavior governed by this fixed point is that of the DP universality class. The explicit 1-loop calculation was necessary to show the stability of this fixed point. Higher orders in perturbation theory modify the two stability exponents quantitatively, and thus affect corrections to scaling. However, they do not change the stability of the fixed point qualitatively since in $\varepsilon$-expansion, $\varepsilon$ is virtually an infinitesimal quantity. This does not answer, as always for the $\varepsilon$-expansion, the question about a possible dimension $d<d_c$ at which the stability breaks down.

We conclude this subsection by collecting the critical exponents of the CDBP transition. The anomalous scaling dimensions of the invariant fields are defined by
\begin{equation}
\eta=\gamma_{0\ast}+\varepsilon\,,\quad\tilde{\eta}=\gamma_{1\ast}-\varepsilon\,. \label{anDim}
\end{equation}
At the stable fixed point they coincide because $\zeta_{0}=(2\beta_{0}
/u_{0})_{\ast}=(\beta_{u}/u)_{\ast}=0$. Also $\gamma_{2\ast}=0$ for $w_{\ast
}=0$. Thus, the coinciding exponents are the same as for DP, $\eta=\tilde{\eta
}$.  The two eigenvalues $\kappa_{0\ast}=-\gamma_{0\ast}/2=(\varepsilon-\eta)/2$ and
$\kappa_{1\ast}$
define the order parameter and the correlation length exponents of DP, respectively,
\begin{subequations}
\label{krit-Ex}
\begin{align}
\beta/\nu &  =(2-\kappa_{0\ast})=(d+\eta)/2 \,,\label{Beta}\\
1/\nu &  =(2-\kappa_{1\ast})\,.\label{Nu}
\end{align}
$\gamma_{\lambda\ast}$ defines the dynamical exponent
\begin{equation}
z=2+\zeta_{\ast}=2+\eta-\gamma_{\lambda\ast}\,.\label{zet}
\end{equation}
As is customary, we furthermore introduce the exponent
\begin{equation}
\Delta=\beta+(z-\eta)\nu\,,\label{Delta}
\end{equation}
\end{subequations}
which is related to the scaling behavior of $H$.

\subsection{Scaling form of the equation of state and correlation lengths}

\begin{figure}[ptb]
\includegraphics[width=5cm]{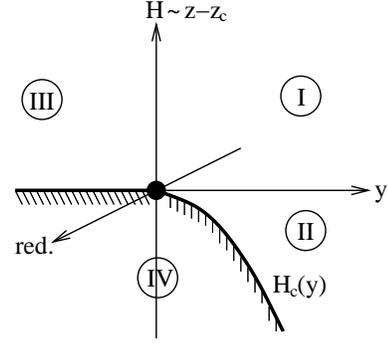}
\caption{Phase diagram with $(y,H)$ as independent variables. The regions I and III are the physical parameter regions for a DP-process. The region III is excluded for the parameters of the statistics of directed branched polymers. The axes marked  \textquotedblleft red.\textquotedblright\ correspond to some redundant variable such as $g_{1}/g_{0}$, $\sigma$, or $c$. The solid dot marks the DP-transition point corresponding to the collapse of large DBP. The line labeled $H_{c}(y)<0$ with $y>0$, where the order parameter $\Phi(z)$ shows a branching point singularity, correspond to large swollen DBP. The shading indicates region IV, where the order parameter $\Phi(z)$ becomes a complex function. At the straight line $H_{c}(y)=0$ with $y<0$ between regions III and IV, the order-parameter shows essential singularities. The lines become surfaces if a redundant variable is admitted.}
\label{fig:phaseDiag}
\end{figure}

Next, we determine the equation of state. To this end, we introduce the dimensionless quantities $\hat{M}=M/\mu^2$, $\hat{H}=H/\mu^4$, and $\hat{y}=y/\mu^2$, and seek $\hat{M}$ as a function of $\hat{H}$ and $\hat{y}$,
\begin{equation}
\hat{M}={F}(\hat{H},\hat{y})\,.\label{Eq-State}
\end{equation}
At the stable DP fixed point, this relation between the three quantities must have scaling form. Using the critical DP exponents, Eq.\ (\ref{krit-Ex}), and exploiting that $\kappa_{01}=\alpha_{00}=0$ at the fixed line $w=0$ which contains the stable fixed point, we obtain from the flow equations (\ref{flow})  the fixed point flow
\begin{subequations}
\label{flow1}
\begin{align}
\ell\frac{d}{d\ell}\hat{y}(\ell)&  =-\frac{1}{\nu}\hat{y}(\ell)\,,\\
\ell\frac{d}{d\ell}\hat{M}(\ell)&  =-\frac{\beta}{\nu}\hat{M}(\ell)\,,\\
\ell\frac{d}{d\ell}\hat{H}(\ell) &  =-\frac{\Delta}{\nu}\hat{H}(\ell) \,
\end{align}
\end{subequations}
with the RG flow-parameter $\ell$ defined by $\mu(\ell)=\ell\mu$.
The scaling form of the equation of state follows now from the solutions of the flow equations, and $\hat{M}(\ell)=F(\hat{H}(\ell),\hat{y}(\ell))$. After elimination of the flow parameter $\ell$, and the redefinitions $\hat{M}\to M$, etc.,  we get the equivalent scaling forms
\begin{equation}
{M}={\vert{H}\vert}^{\beta/\Delta}{\bar f}_{\pm}({y}/{\vert{H}\vert}^{1/\Delta})
=y^{\beta}{f}(H/y^{\Delta})\,,\label{Eq-State1}
\end{equation}
where the subscript $\pm$ indicates the sign of $H$. Note that this field theoretic result  is fully consistent with the mean-field equation of state as stated in Eq.~(\ref{SPL}).

Written in its scaling forms~(\ref{Eq-State1}), the equation of state leads to the following observations, see Fig.\ (\ref{fig:phaseDiag}):  the region of the phase diagram corresponding to DP is given by $M\geq 0$ and $H\geq 0$, where the primary control parameter $y$ can take both signs \cite{JaKuOe99}. Note, however, that the stability condition, Eq.\ (\ref{StabBdg}) restricts the variable $y$ in our polymer problem to $y\geq0$, and $y=0$ at the collapse transition, coinciding with the DP transition point. The primary control parameter here is the external field $H\sim z-z_{c}$. The DP transition point becomes now a tricritical point given by $H_{c}=0$ and $y_{c}=0$. The control parameter $y$ constitutes here the crossover variable which drives the critical behavior from the DP-universality class to the critical point of the dynamical Yang-Lee model that we have discussed in Sec.~\ref{sec:nonCollapsing}. Hence, for $y>0$ we expect a critical line $H=H_c(y)<0$ with $M<0$ in a region of the phase diagram not accessible for DP itself.

To discuss the crossover in some more detail, we note that in mean-field theory (where  $\beta=1$ and $\Delta=2$), the scaling functions defined by Eq.~(\ref{Eq-State1}) are simply given by
\begin{subequations}
\label{Eq-State-mf}
\begin{align}
{f}(x)&=\sqrt{x+1}-1\,,\label{Eq-State-mf1}
\\
\bar{f}_{\pm}(\bar{x})&=\sqrt{{\bar x}^2\pm 1}-\bar x\,.\label{Eq-State-mf2}
\end{align}
\end{subequations}
These equations show the tricritical point at $x=0$ and $\bar{x}=\infty$. The critical line is described by the branchpoint singularity at $x=-1$ or $\bar{x}=1$, respectively. However, mean-field theory for the crossover is applicable only for $d>6$, where both types of polymers, collapsing and swollen, have mean-field behavior.

To determine the general form of the crossover, we assume as usual that $f(x)$ has in either case a singularity at same negative value $x_{\ast}$ of $x$ \cite{LaSa84}. To be specific, we assume that
\begin{equation}
f(x)=f_0(x)+(x-x_{\ast})^{\dot{\theta}-1}f_1(x)\,.
\end{equation}
$f_0(x)$ is the analytic part of $f(x)$ at $x=x_{\ast}$, in particular $f_0(x_{\ast})=-A<0$.  $f_1(x)$ may havefurther singularities at $x=x_{\ast}$ but it is assumed that $f_1(x_{\ast})=B$ exists. $\dot{\theta}$ denotes the entropic exponent of the swollen polymers, cf.\ Sec.~\ref{sec:nonCollapsing}. With these assumptions, we find in  the vicinity of the singular curve $H=H_c(y)=x_{\ast}y^{\Delta}$ that
\begin{equation}
M\approx -Ay^{\beta}+By^{(\theta-\dot{\theta})\Delta} (H-x_{\ast}y^{\Delta})^{\dot{\theta}-1}\,,
\end{equation}
where
\begin{equation}
\label{thetaDef}
\theta=1+\beta/\Delta
\end{equation}
is the entropic exponent of the collapsing polymers. To conclude our discussion of the crossover, we emphasize that the problem involves \emph{two distinct kinds of nontrivial critical behavior with different upper critical dimensions}. Thus, one cannot simply extract via analytical continuation $\varepsilon$-expansion results for $f(x)$ from the known results for the DP equation of state~\cite{JaKuOe99}. A glance at the two-loop part of the latter shows, in deed, non-summable singularities at $x=-1$.

Now, we turn to the scaling form of the correlation lengths, i.e., the diverging length-scales in transversal and longitudinal direction. To this end, we consider the scaling behavior of the correlation functions $G_{k,\tilde{k}}$ of $k$ fields $\phi$ and $\tilde{k}$ fields $\tilde{\phi}$. The RGE for these correlation functions follows readily from the general RGE, see Eqs.~(\ref{inv-RGE-Diff}) and (\ref{inv-RG-Feld}).
At the stable DP fixed point with $\gamma_{0\ast
}=\gamma_{1\ast}=\gamma_{\ast}=\eta$, $\gamma_{2\ast}=0$, this RGE reads
\begin{align}
\label{RGEcorrFkt}
\left[  \mathcal{D}_{\mu}+\frac{\eta}{2}(k+\tilde{k})\right]  G_{k,\tilde{k}
}(\{\mathbf{x},t\},y,H)=0\,.
\end{align}
Via solving RGE~(\ref{RGEcorrFkt}), we obtain the scaling forms
\begin{align}
&G_{k,\tilde{k}} (\{\mathbf{x},t\},y,H)
\nonumber\\
&=\left\vert H\right\vert ^{(k+\tilde{k})\beta/\Delta}
{\bar{f}}_{\pm;k,\tilde{k}}(\{\mathbf{r/}\xi_{\perp},t/\xi_{\Vert}\},
y/\left\vert H\right\vert ^{1/\Delta})\nonumber\\
&=y^{(k+\tilde{k})\beta}{f}_{k,\tilde{k}}(\{\mathbf{r/}\xi_{\perp},t/\xi_{\Vert}\},H/y ^{\Delta})
\end{align}
of the correlation functions, where ${f}_{\pm;k,\tilde{k}}$ and ${\bar{f}}_{k,\tilde{k}}$ are scaling functions with the subscript $\pm$ denoting again the sign of $H$. Note that we need $H\leq 0$, for $H<0$ outside the physical region of DP. This scaling form implies the following scaling forms for the correlation lengths in the transversal and longitudinal directions,
\begin{subequations}
\label{length-scales}
\begin{align}
\xi_{\perp}  &  =\left\vert{z-z_{c}}\right\vert^{-\nu/\Delta}
\hat{f}_{\bot}\left(y/\left\vert{z-z_{c}}\right\vert^{1/\Delta}\right)
\,,\label{length-senk}\\
\xi_{\Vert}  &  =\left\vert{z-z_{c}}\right\vert^{-z\nu/\Delta}
\hat{f}_{\Vert}\left(  y/\left\vert{z-z_{c}}\right\vert^{1/\Delta}\right)
\,, \label{length-parall}
\end{align}
\end{subequations}
with scaling functions $\hat{f}_{\bot}$ and $\hat{f}_{\Vert}$ with an analogous crossover behavior as discussed for the equation of state.

\section{Asymptotic properties of collapsing directed branched polymers}
\label{sec:polymerScaling}

Our field theory of Sec~\ref{sec:fieldTheory} provided us with a number of results, in particular the scaling form of the equation of state and the correlation lengths. Though certainly interesting in their own right as they stand, they are not yet in a form or language that is customary in polymer theory. Here, we translate our results into results for the animal numbers and the radii of gyration.

As discussed in Sec.~\ref{sec:towardsFieldTheory}, the number of different polymer configurations
$\mathcal{A}(N)$ for large $N$ is up to an nonuniversal exponential factor $\mu_{0}^{N}$ proportional to  $\mathcal{P}(N)$, and the relation of the latter to $\Phi(z)$ via inverse Laplace led asymptotically to Eq.\ (\ref{Probab_N}). Now we recall that $\Phi(z)$ is linearly related to $M$, and revisit our result (\ref{Eq-State1}) for the equation of state which we rewrite as
\begin{equation}
\Phi(z)-\Phi(z_{c}) \sim \left\vert{z-z_{c}}\right\vert^{\beta/\Delta}
\hat{f}\left(y/\left\vert{z-z_{c}}\right\vert^{1/\Delta}\right)  \,.
\end{equation}
Inserting this into the inverse Laplace transformation~(\ref{Probab_N}), we obtain for $N\gg1$ the following scaling form for the animal numbers:
\begin{equation}
\mathcal{A}(N,y)\sim N^{-\theta}\,f\left(yN^{\phi}\right)\mu_{0}^{N}
\end{equation}
with the animal exponent $\theta$ which is the same as the entropic exponent defined in Eq.~(\ref{thetaDef}), and the crossover exponent
\begin{equation}
\phi=1/\Delta\,.
\end{equation}

We see here at the instance of the animal numbers, that the transcription of the field theoretic scaling result to the corresponding scaling result in the polymer language asymptotically for $N\gg 1$ via inverse Laplace transformation amounts to simply replacing $\left\vert{z-z_{c}}\right\vert$ by $N^{-1}$, see also the second part of Eq.\ (\ref{Probab_N}). In this way, after the inverse Laplace transformation of the correlation functions, space- and time-coordinates, which originally scale with the correlation lengths, now scale with the appropriate powers of $N$. Consequently, we obtain for the longitudinal and the transversal radii of gyration as counterparts of the correlation lenghts
\begin{subequations}
\begin{align}
R_{\parallel}(N,y)&\sim N^{\nu_{A,\Vert}}\,f_{\parallel} \left(  yN^{\phi}\right) \,,
\\
R_{\bot}(N,y)&\sim N^{\nu_{A,\bot}}\,f_{\bot} \left(  yN^{\phi}\right) \,,
\end{align}
\end{subequations}
with critical exponents
\begin{subequations}
\begin{align}
\nu_{A,\Vert}  &  =z\nu/\Delta\,,\\
\nu_{A,\bot}  &  =\nu/\Delta\,,
\end{align}
\end{subequations}
where $z$ is once again the dynamical exponent of DP which should not de confused with the variable $z$ introduced through the Laplace transformation.

It remains to state the $\varepsilon$-expansions results for the polymer exponents encountered in this section. Using the relations of these exponents to the DP-exponents, which are  known to second order in $\varepsilon$~\cite{Ja81,Ja01,JaTa05}, we obtain
\begin{subequations}
\begin{align}
\theta &  =\frac{3}{2}-\frac{\varepsilon}{12}\Big[1+\Big(\frac{37}{288}%
+\frac{53}{144}\ln(4/3)\Big)\varepsilon+\ldots\Big]\,,\\
\phi &  =\frac{1}{2}-\frac{\varepsilon^{2}}{18}\Big[1+\ldots\Big]\,,\\
\nu_{A,\Vert}  &  =\frac{1}{2}+\frac{\varepsilon}{24}\Big[1+\Big(\frac
{13}{288}-\frac{55}{144}\ln(4/3)\Big)\varepsilon+\ldots\Big]\,,\\
\nu_{A,\bot}  &  =\frac{1}{4}+\frac{\varepsilon}{32}\Big[1+\Big(\frac{43}%
{288}-\frac{17}{144}\ln(4/3)\Big)\varepsilon+\ldots\Big]\,.
\end{align}
\end{subequations}
In table~\ref{table:CDBPexponents}, we compare these $\varepsilon$-expansions to results
of numerical simulations as listed, e.g., in the review-article of
Hinrichsen \cite{Hin01} or the recent book of Henkel, Hinrichsen, and
L\"{u}beck \cite{HeHiLu08}, which, of course, are obtained by simulations of the DP-transition, and not for the polymer problem itself. Down to $d=2$, the agreement between field theory and simulations is remarkably good.
\begin{table}
\begin{tabular}
[c]{l|llll}%
$d:\quad$ & $\quad1\qquad$ & $\;\;2\qquad$ & $\;\;3\qquad$ & $\;\;4\qquad
$\\\hline\hline
$\theta:\quad$ & $\;1.10825\quad$ & $1.27$ & $1.40$ & $1.5$\\
$(\varepsilon-Exp.)\;$ & $\;1.074$ & $1.255$ & $1.397$ & $3/2$\\\hline
$\phi:\quad$ & $\;0.39151\quad$ & $0.459$ & $0.49$ & $0.5$\\
$(\varepsilon-Exp.)\;$ & $\;0.375$ & $0.444$ & $0.486$ & $1/2$\\\hline
$\nu_{\parallel}:\quad$ & $\;0.69007$ & $0.584$ & $0.54$ & $0.5$\\
$(\varepsilon-Exp.)\;$ & $\;0.601$ & $0.573$ & $0.539$ & $1/2$\\\hline
$\nu_{\bot}:\quad$ & $\;0.451494$ & $0.337$ & $0.285$ & $0.25$\\
$(\varepsilon-Exp.)\;$ & $\;0.376$ & $0.327$ & $0.285$ & $1/4$%
\end{tabular}
\caption{Values of the CDBP exponents for various transversal dimensions $d$ as produced by our field theory and numerical simulations~\cite{Hin01,HeHiLu08}.}
\label{table:CDBPexponents}
\end{table}

\section{Concluding remarks}
\label{sec:concludingRemarks}
In summary, we have developed a model for directed branched polymers in the framework of dynamical field field theory. As far as mean-field theory was concerned, our model reproduced a number of well known results and, therefore, passed important consistency checks. The same holds true for the field theory of swollen directed branched polymers which we extracted from our general model. Our main focus laid on the collapse transition of directed branched polymers for which, to our knowledge, no field theory existed hitherto. We showed to arbitrary order in perturbation theory that this transition belongs to the DP universality class. Because we used $\varepsilon$-expansion, we can be sure about the stability of the DP fixed-point only for those dimensions for which the $\varepsilon$-expansion is valid. Given that it is known from earlier work that the collapse transition of directed branched polymers belongs to the DP universality class in $1+1$-dimensions and that our present work shows that this is also the case in the vicinity of the upper critical dimension $4+1$, we are led to expect that this holds true in any dimension. We calculated the scaling behavior of several quantities of potential experimental relevance such as gyration radii and the probability distribution for finding a polymer consisting of $N$ monomers.

\begin{acknowledgments}
  This work was supported in part (O.S.) by the National Science Foundation under
  grant No.\ DMR 0804900.
\end{acknowledgments}

\appendix

\section{1-loop calculation}
\label{app:1Loop}
\begin{figure}[ptb]
\includegraphics[width=8.0cm]{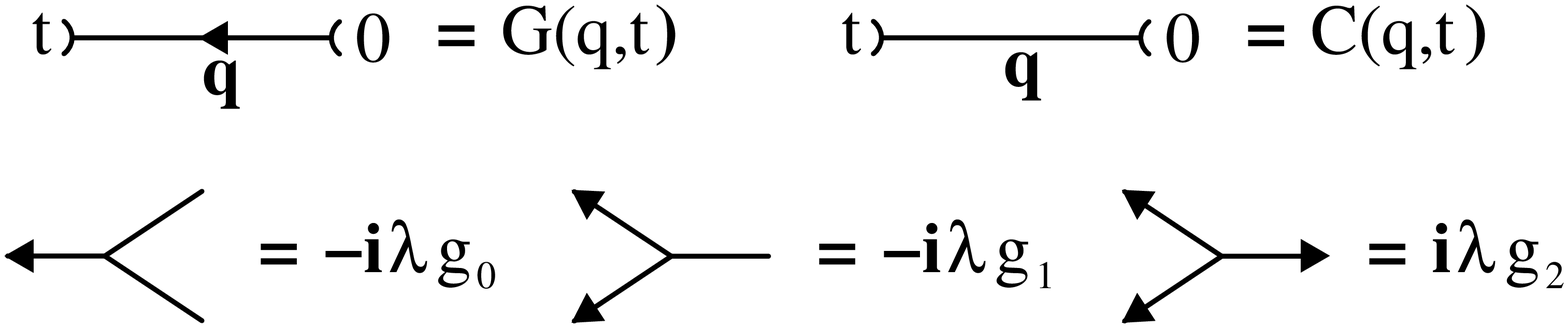}
\caption{Diagrammatic elements.}
\label{fig:elemente}
\end{figure}

\begin{figure}[ptb]
\includegraphics[width=8.6cm]{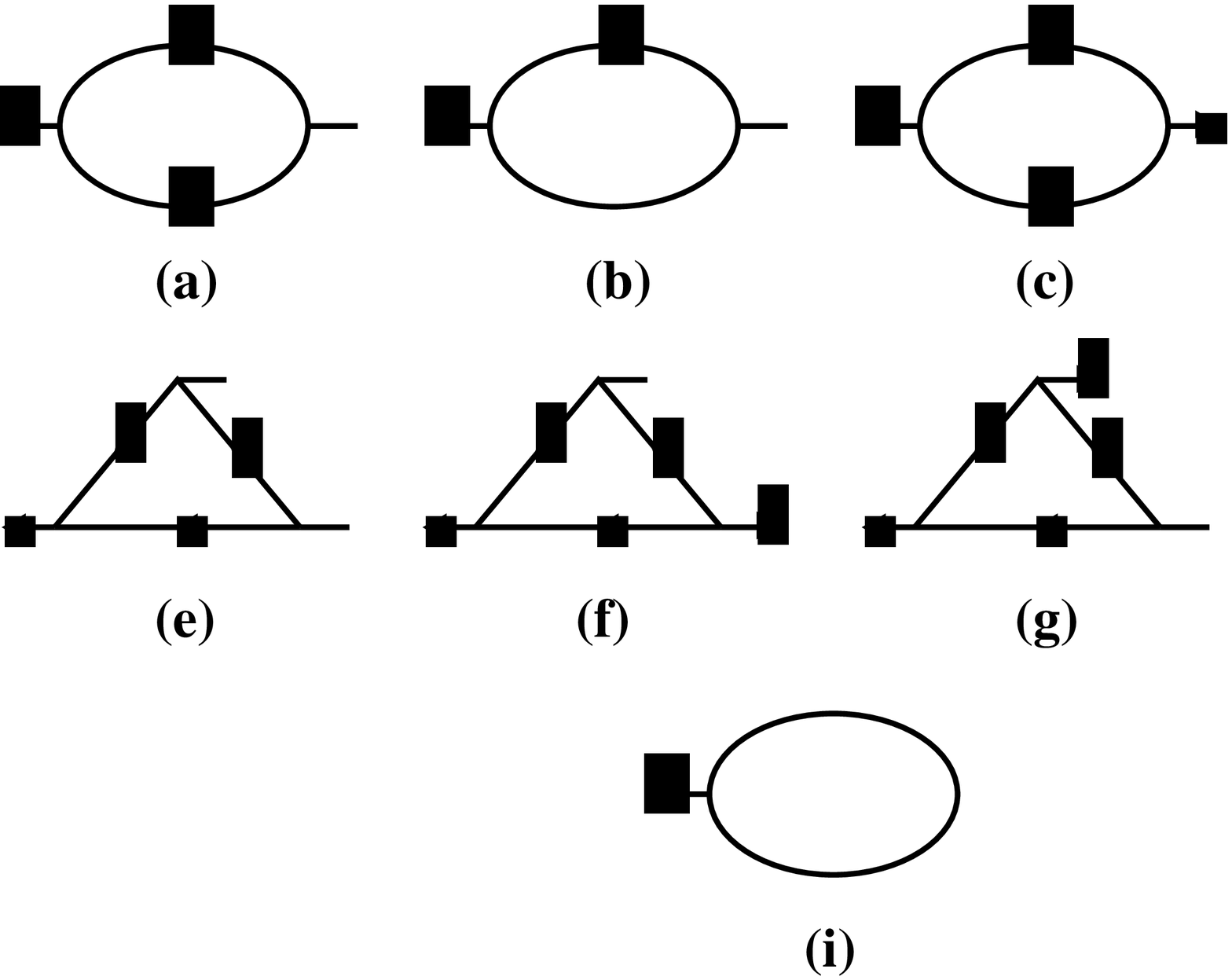}
\caption{1-loop contributions to the superficially divergent vertex functions.}
\label{fig:1-loop}
\end{figure}

In this appendix we provide some details of our 1-loop calculation. The diagrammatic elements entering this calculation are the Gaussian propagator
\begin{align}
G (\brm{q}, t) = \Theta (t) \, \exp \left[  - \lambda \left(  \brm{q}^2 + \hat{\tau}_0 \right) t \right] ,
\end{align}
the correlator
\begin{align}
C (\brm{q}, t) = \frac{\hat{\tau}_1/2}{\brm{q}^2 + \hat{\tau}_0}\, \exp \left[  - \lambda \left(  \brm{q}^2 + \hat{\tau}_0 \right) |t|\right] ,
\end{align}
and the three-leg vertices $-i \lambda g_0$, $-i \lambda g_1$ and $i \lambda g_2$, see Fig.~\ref{fig:elemente}.
As usual, we concentrate on those vertex functions that are superficially divergent in the ultraviolet. These are $\Gamma_{1,1}$, $\Gamma_{2,0}$, $\Gamma_{1,2}$, $\Gamma_{2,1}$, $\Gamma_{3,0}$ and $\Gamma_{1,0}$. The diagrammatic contributions to these vertex functions to 1-loop order are collected in Fig.~\ref{fig:1-loop}.

To give an example for the calculation of the various 1-loop contributions, let us consider the diagram 3(c). The mathematical expression for this diagram is
\begin{align}
3(c) &= \frac{1}{2} \, (- i \lambda g_0)  (i \lambda g_2)\int_0^\infty dt \, \exp \left[  - \left( i \omega + 2 \lambda \hat{\tau}_0 \right)  t \right]
\nonumber \\
& \times \frac{1}{(2\pi)^d} \int d^d k \,  \exp \left[  - \lambda \left(  \brm{k}^2 + (\brm{q} + \brm{k} )^2 \right) t \right]
\nonumber\\
&+ (\omega \to - \omega),
\end{align}
where the last term indicates that this diagram can also be drawn with the arrows on the propagators pointing to the right which corresponds to the first term with $\omega$ replaced by $-\omega$. The momentum and time integrations are straightforward. They lead to
\begin{align}
3(c) &= - \frac{\lambda g_0 g_2}{4} \, \frac{\Gamma (\varepsilon/2 -1)}{(4 \pi)^{d/2}} \, \left[  \frac{i \omega}{2\lambda} + \hat{\tau}_0 + \frac{\brm{q}^2}{4} \right]^{1-\varepsilon/2}
\nonumber\\
&+ (\omega \to - \omega).
\end{align}
$\varepsilon$-expansion finally produces
\begin{align}
3(c) &= -  \lambda g_0 g_2\frac{G_\varepsilon}{\varepsilon} \, \left[  \hat{\tau}_0 + \frac{\brm{q}^2}{4} \right] ,
\end{align}
where we have omitted finite contributions that are inconsequential for our quest.
Note that the contributions proportional to $\omega$ cancel as it should be the case. Note also that this diagram contributes a term proportional to $\brm{q}^2$ to $\Gamma_{2,0}$. A similar situation occurs for the collapse transition of isotropic randomly branched polymers. This and the renormalization mandated by it was overlooked in previous work \cite{LuIs78,HaLu81} and hence led to incorrect results.

The remaining diagrams in Fig.~\ref{fig:1-loop} can be calculated by similar means. To save space, we merely list here the so-obtained results for the as yet unrenormalized vertex functions:
\begin{subequations}
\begin{align}
\mathring\Gamma_{1,1} &=i \omega + \lambda \left( \hat{\tau}_0 + \brm{q}^2 \right)- 3(a) -3(b) \nonumber\\
&= i \omega + \lambda \left( \hat{\tau}_0 + \brm{q}^2 \right) \nonumber \\
&+ \lambda \hat{\tau}_0^{-\varepsilon/2} \, \frac{G_\varepsilon}{2\varepsilon} \,  g_0 \bigg\{  g_0 \hat{\tau}_1
 - g_1\left(\hat{\tau}_0 +  \frac{i \omega}{2\lambda} +  \frac{\brm{q}^2}{4} \right) \bigg\},
\\
\mathring\Gamma_{2,0} &=- \lambda \hat{\tau}_1 -3(c) -3(d) \nonumber\\
&= - \lambda \hat{\tau}_1+ \lambda \hat{\tau}_0^{-\varepsilon/2} \, \frac{G_\varepsilon}{\varepsilon} \,  g_0 \bigg\{   g_1 \hat{\tau}_1 +  g_2 \left(\hat{\tau}_0 +\frac{\brm{q}^2}{4}\right)  \bigg\},
\\
\mathring\Gamma_{1,2} & = i \lambda g_0 -3(e) =i \lambda g_0- i \lambda \hat{\tau}_0^{-\varepsilon/2} \, \frac{G_\varepsilon}{\varepsilon} \, g_0^2 g_1 \, ,
\\
\mathring\Gamma_{2,1} & =  i \lambda g_1 -3(f) -3(g) \nonumber\\
&= i \lambda g_1 + i \lambda \hat{\tau}_0^{-\varepsilon/2} \, \frac{G_\varepsilon}{\varepsilon} \, g_0 \left\{ g_0 g_2   - g_1^2\right\} ,
\\
\mathring\Gamma_{3,0} & =- i \lambda g_2 -3(h) =- i \lambda g_2 + i \lambda \hat{\tau}_0^{-\varepsilon/2} \, \frac{3 G_\varepsilon}{\varepsilon} \, g_0 g_1 g_2 \, ,
\\
\mathring\Gamma_{1,0} & =i \lambda \hat{h} -3(i) = i \lambda \hat{h} - i \lambda \hat{\tau}_0^{-\varepsilon/2} \, \frac{G_\varepsilon}{2\varepsilon} \, g_0 \hat{\tau}_0 \hat{\tau}_1\, .
\end{align}
\end{subequations}
It can easily be checked that these results satisfy the Ward identities discussed in Sec.~\ref{subsec:Wardetc}. Application of the renormalization scheme~(\ref{Ren}) leads to renormalized vertex functions
\begin{subequations}
\begin{align}
\Gamma_{1,0} & =Z^{1/2}\mathring\Gamma_{1,0}\,,\quad \Gamma_{1,1} =Z \mathring\Gamma_{1,1}\,,\\
\Gamma_{2,0} &=Z\left(\mathring\Gamma_{2,0} +2K \mathring\Gamma_{1,1}\right)\,,\quad
\Gamma_{1,2} =Z^{3/2}\mathring\Gamma_{1,2}\,,\\
\Gamma_{2,1} &=Z^{3/2}\left(\mathring\Gamma_{2,1} +2K \mathring\Gamma_{1,2}\right)\,,\\
\Gamma_{3,0} &=Z^{3/2}\left(\mathring\Gamma_{3,0} +3K \mathring\Gamma_{2,1} + 3K^2 \mathring\Gamma_{1,2}\right)\,,
\end{align}
\end{subequations}
and then finally to the 1-loop results for the renormalization factors and additive counter-terms stated in Eq.~(\ref{Ren-1L}).

\section{Essential singularities}
\label{app:essentialSingularities}
In this appendix, we briefly discuss essential singularities appearing in the generating function $\Phi(z)$ for $y<0$ while crossing the separation line between regions III and IV in the phase diagram, Fig.~\ref{fig:phaseDiag}. These singularities produced by instantons are intimately related to the well known so-called false-vacuum problem \cite{Col77}. They lead to the behavior
\begin{equation}
\mathcal{P} \sim N^{-\theta'}\exp{\bigl[-b(|y|^{1/\phi}N)^{\zeta}}\bigr]
\label{essP}
\end{equation}
of the distribution function $\mathcal{P}$ with $\zeta < 1$ and a nonuniversal amplitude $b$.  Our discussion follows that of Harris and Lubensky \cite{HaLu81}, and Lubensky and McKane \cite{LuMcK81} which are based on Langers classical paper \cite{La67} on the droplet model.

\begin{figure}[ptb]
\includegraphics[width=5.5cm]{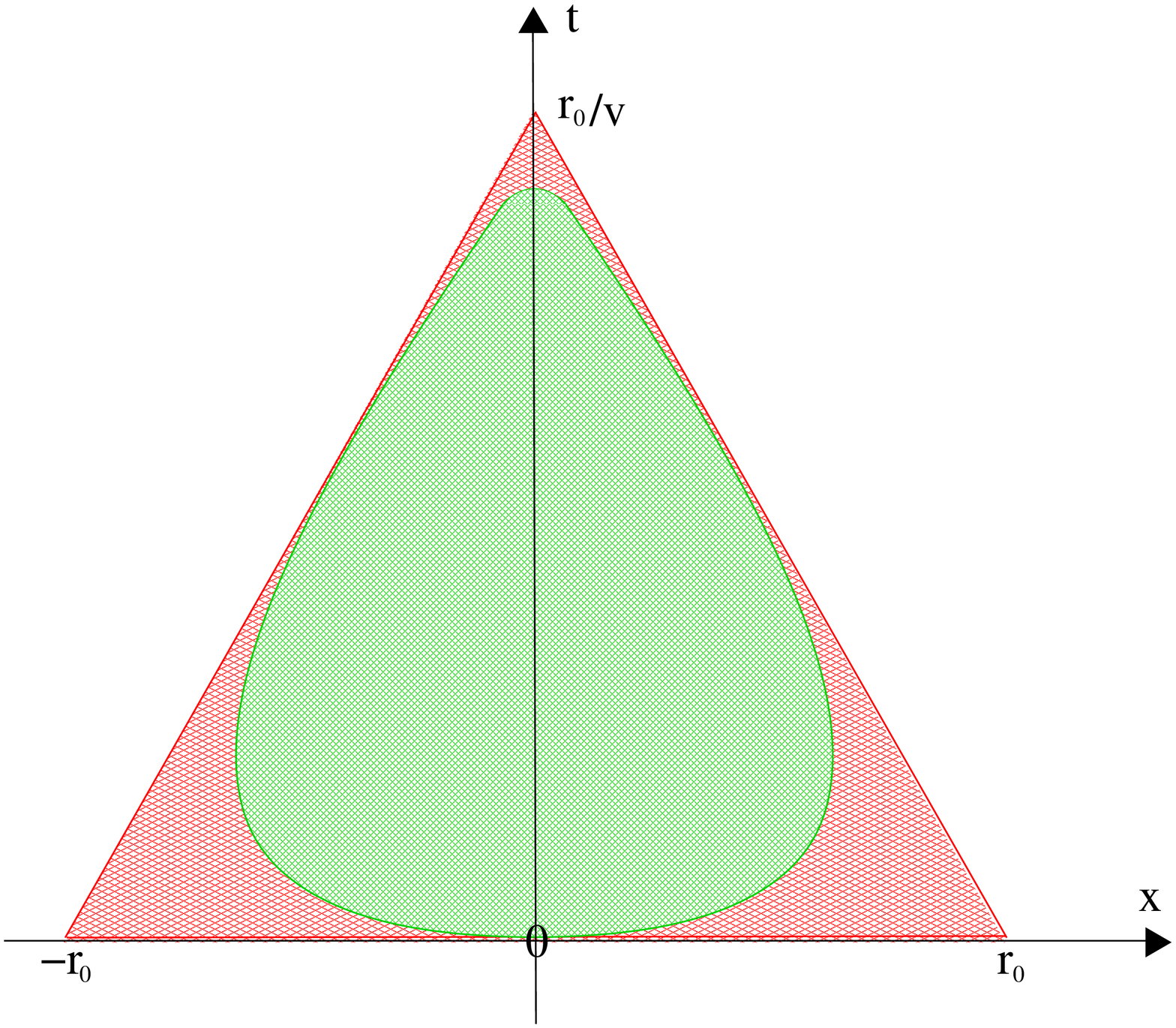}
\caption{(Color online) $(d+1)$-dimensional cut of a droplet configuration (green inner part) of the true vacuum ($s=0$, $\tilde{s}=-2|\tau|/g$) in the sea of the false vacuum ($s=2|\tau|/g$, $\tilde{s}=0$) and its approximation for $r_0\gg 1/\sqrt{|\tau|}$ by a $(d+1)$-dimensional cone (green inner and red outer part).}
\label{fig:droplet}
\end{figure}

We have shown in the main text that the universal critical behavior of collapsing directed branched polymers is equivalent to that of DP. Being interested in essential singularities, we are hence led here to studying the behavior of the cluster distribution function of finite DP-clusters in the active phase above the transition point (where also an infinite cluster in time and space is generated with finite probability). As the basis for this study, we use the stochastic functional (\ref{J_ende}) in its usual symmetrized real form with $g_0=g_1=g$, $g_2=0$, and with $\hat\tau_1$ eliminated. We set $\hat\tau_0=\tau$ and work with
\begin{equation}
\mathcal{J} =\int d^{d}xdt\,\lambda\Big\{\tilde
{s}\Big[\lambda^{-1}\partial_{t}+\tau-\nabla^{2}+\frac{g}{2}\bigl(s-\tilde{s}\bigr)\Big]s
-{h}\tilde{s}\Big\}\,. \label{J_DP}\\
\end{equation}
We need to understand the behavior of functional integrals with weight $\exp(-\mathcal{J}[s,\tilde{s}])$ for $\tau=-|\tau|$ near the first order line $h\approx0$. To this end, we seek for saddle points. If $h>0$ near zero, the true vacuum, i.e., the saddle point producing the lowest value of $\mathcal{J}$, is clearly the mean-field solution ($s=2|\tau|/g$, $\tilde{s}=0$), up to corrections proportional to $h$. However, if we cross the border line at $h=0$ and go to  small $h<0$, the mean field solution becomes a false vacuum. At first sight, the saddle point solution ($s=0$, $\tilde{s}=-2|\tau|/g$) appears to be the true vacuum. However, the response field has to satisfy the boundary condition $\tilde{s}\to 0$ when the space coordinate or time tend to infinity. Thus, only non-stationary inhomogeneous solutions of the saddle-point equations, i.e., instantons or droplets, are allowed, see Fig.~(\ref{fig:droplet}).

To search for droplet solutions, it is useful to switch to dimensionless quantities. We introduce new fields $s=2|\tau|\varphi/g$, $\tilde{s}=2|\tau|\tilde{\varphi}/g$, $h=4\tau^2 k/g$, and scale time and space by $\lambda |\tau|t=t'$, $|\tau|^{1/2}\underline{x}=\underline{x}'$. In the following we will drop the primes for notational simplicity. The stochastic functional becomes
\begin{subequations}
\label{action}
\begin{align}
\mathcal{J} &= \frac{4}{g^2} |\tau|^{2-d/2}\mathcal{A}\,,
\\
\label{actionA}
\mathcal{A} &= \int d^d x dt \Big\{\tilde{\varphi}\dot{\varphi}-H(\tilde{\varphi},\varphi)\Big\}\,,\\
H(\tilde{\varphi},\varphi) &= -\nabla\tilde{\varphi}\nabla{\varphi}+\tilde{\varphi}\big(1-\varphi+\tilde{\varphi}\big)\varphi
+k\tilde{\varphi}\,.
\end{align}
\end{subequations}

\begin{figure}[ptb]
\includegraphics[width=3.5cm]{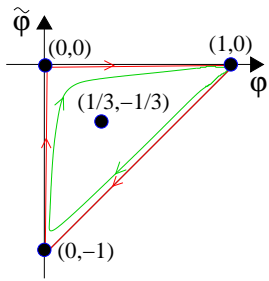}
\caption{(Color online) The orbit of the droplet's instanton-configuration with $|\underline{x}|=\textit{const.}$ between the false and the true vacuum (green inner orbit), and its approximation by kink-solutions (red outer orbit).}
\label{fig:instanton}
\end{figure}

The functional integral over $\exp(-\mathcal{J}[s,\tilde{s}])$ is definitely equal to one. However, if we add a source term $qs(\underline{x}_0,t_0)$ to the exponent, the saddle-point equations
\begin{subequations}
\label{saddelpoint}
\begin{align}
\dot{\varphi} &= \nabla^2{\varphi}+\big(1-\varphi+2\tilde{\varphi}\big)\varphi +k\,,
\\
-\dot{\tilde\varphi} &=\nabla^2{\tilde\varphi}+\big(1+\tilde\varphi-2{\varphi}\big)\tilde\varphi\, ,
\end{align}
\end{subequations}
generate a droplet of the true vacuum that has its tip at $(\underline{x}_0,t_0)$ and faces backward in time, see Fig.~(\ref{fig:droplet}). We are interested in the critical droplet configuration whose lateral extension diverges as $r = r_c \sim |k|^{-1}$ with $|k|\to 0$. We approximate this configuration by a critical spherical symmetric cone, see Fig.~(\ref{fig:droplet}). Hence, we look for solitary waves $\varphi(\rho)$ and $\tilde{\varphi}(\rho)$ with $\rho = r+vt-r_c$ as kink-solutions of the saddle-point equations (\ref{saddelpoint}) which have their variations near $\rho =0$. By neglecting $k$ as well as $1/|\underline{x}|$ in comparison to $v$, we obtain from Eqs.~(\ref{saddelpoint}) the differential equations
\begin{subequations}
\begin{align}
v{\varphi}' &= {\varphi}''+\big(1-\varphi+2\tilde{\varphi}\big)\varphi \,,
\\
-v{\tilde{\varphi}}' &={\tilde\varphi}''+\big(1+\tilde\varphi-2{\varphi}\big)\tilde\varphi\,,
\label{saddelpoint-app}
\end{align}
\end{subequations}
where the stroke means differentiation with respect to $\rho$. The properties of the kink-solutions, see Fig.~(\ref{fig:instanton}), and the fluctuations around them are extensively discussed in the work of Alessandrini \emph{et al.}\ \cite{AlAmCi77} on classical kinks and their quantization in RFT. Also, in this work the appearance of the critical velocities $v = -\infty$ (the base of the cone) and $v = 2$ (the envelope of the cone) is shown.

Taking all things together, the action $\mathcal{A}$, Eq.~(\ref{actionA}), of the large droplet consists of the usual terms arising from the volume $\sim r^{d+1}$ of the droplet times $k\tilde{\varphi}$, where $\tilde{\varphi}$ takes its value in the true vacuum, namely $-1$, and a surface part $\sim r^d$ arising from the variations of the kink-solutions (in the limit $v \to 2$ this surface part comes only from the base of the cone). We get, therefore,
\begin{equation}
\mathcal{A} = -a|k|r^{d+1} + br^{d}\,,
\end{equation}
where $a$ and $b$ are positive constants. Minimizing $\mathcal{A}$ with respect to $r$, we obtain $r = r_c\sim 1/|k|$, and finally
\begin{equation}
\mathcal{J}_c =\frac{4}{g^2} |\tau|^{2-d/2}\mathcal{A}_c \sim |h|^{-d}\,.
\label{critJ}
\end{equation}

Considering the eigenvalues of the Gaussian fluctuations about the critical instanton, we have (i) one negative eigenvalue coming from the uniform variations of the extension $r$ of the critical droplet leading to the imaginary part of $\Phi$, (ii) $(d+1)$ eigenvalues $0$ coming from the translations in space and time of the droplet and their compensation by the pinning of the droplet at its source, and (iii) a series of positive eigenvalues coming from the Goldstone modes of the broken translation symmetries. For fully spherical symmetric droplets, the calculation of the latter eigenvalues is straightforward~\cite{GuNiWa80}. In the present problem, however, the droplets are conical (spherical symmetric in space but directed in time). For this configuration, the Goldstone-mode eigenvalues have not yet been calculated, as far as we know, and their calculation represents an interesting and challenging opportunity for future work.

Finally, we return to the cluster distribution function $\mathcal{P}$, Eq.~(\ref{essP}), which follows via inverse Laplace transformation as discussed extensively in the main text. For the exponent $\zeta$, this leads to $\zeta = d/(d+1)$. To determine $\theta'$, however, we first need to know the Goldstone-mode eigenvalues touched upon in the preceding paragraph. Thus, we here have to leave the interesting problem of calculating $\theta'$ for future work.

\end{document}